\documentclass[fleqn,usenatbib]{mnras}

\usepackage{newtxtext,newtxmath}

\usepackage[T1]{fontenc}
\usepackage{ae,aecompl}

\usepackage{graphicx}	
\usepackage{amsmath}	
\usepackage{subcaption}
\usepackage[export]{adjustbox}
\usepackage{makecell}
\usepackage{siunitx}

\title[\textit{TESS} Observations of Core-Collapse Supernovae]{High-Cadence, Early-Time Observations of Core-Collapse Supernovae From the \textit{TESS} Prime Mission}

\author[Vallely et al.]
{P. J. Vallely$^{1}$\thanks{E-mail: vallely.7@osu.edu (PJV)},
C. S. Kochanek$^{1,2}$,
K. Z. Stanek$^{1,2}$,
M. Fausnaugh$^{3,4}$,
\newauthor
B. J. Shappee$^{5}$
\\
$^{1}$Department of Astronomy, The Ohio State University, 140 West 18th Avenue, Columbus, OH 43210, USA\\
$^{2}$Center for Cosmology and AstroParticle Physics, The Ohio State University, 191 W. Woodruff Ave., Columbus, OH 43210, USA\\
$^{3}$Department of Physics, Massachusetts Institute of Technology, Cambridge, MA 02139, USA\\
$^{4}$Kavli Institute for Astrophysics and Space Research, Massachusetts Institute of Technology, Cambridge, MA 02139, USA\\
$^{5}$Institute for Astronomy, University of Hawai'i, 2680 Woodlawn Drive, Honolulu, HI 96822, USA \\
}

\date{Accepted XXX. Received YYY; in original form ZZZ}

\pubyear{2020}

\begin{document}
\label{firstpage}
\pagerange{\pageref{firstpage}--\pageref{lastpage}}
\maketitle

\begin{abstract}
We present observations from the Transiting Exoplanet Survey Satellite (\textit{TESS}) of twenty bright core-collapse supernovae with peak \textit{TESS}-band magnitudes $\lesssim18$ mag.
We reduce this data with an implementation of the image subtraction pipeline used by the All-Sky Automated Survey for Supernovae (ASAS-SN) optimized for use with the \textit{TESS} images.
In empirical fits to the rising light curves, we do not find strong correlations between the fit parameters and the peak luminosity.
Existing semi-analytic models fit the light curves of the Type II supernovae well, but do not yield reasonable estimates of the progenitor radius or explosion energy, likely because they are derived for use with ultraviolet observations while \textit{TESS} observes in the near-infrared.
If we instead fit the data with numerically simulated light curves, the rising light curves of the Type~II supernovae are consistent with the explosions of red supergiants.
While we do not identify shock breakout emission for any individual event, when we combine the fit residuals of the Type II supernovae in our sample, we do find a $>5\sigma$ flux excess in the $\sim 0.5$~day before the start of the light curve rise.
It is likely that this excess is due to shock breakout emission, and that during its extended mission \textit{TESS} will observe a Type II supernova bright enough for this signal to be detected directly.
\end{abstract}




\section{Introduction}
\label{sec:intro}

Core-collapse supernovae are a broad class of astronomical transients produced by the explosion of high mass $(\gtrsim 8 M_\odot)$ stars at the end of their lives.
The most common of these core-collapse subtypes, indeed the most common of all supernovae subtypes in a volume-limited sample, are the Type II supernovae \citep[SNe~II;][]{2011Li,2009Smarttetal,2019HoloienaCat}.
The progenitors of these explosions are red supergiants with massive hydrogen envelopes that extend out to hundreds of solar radii (e.g., \citealt{2009Smartt,2015Smartt}), and they are empirically defined by the presence of hydrogen emission lines in their spectra \citep{1997Filippenko}.
Some SNe~II show evidence of interaction with dense circumstellar material (CSM) in the form of narrow emission features, leading to their designation as SNe~IIn \citep{1990Schlegel,1994Chugai}.

While the majority of supernovae in a volume-limited sample will be Type II SNe, about one-fifth of the sample will be comprised of Type Ib and Ic events \citep[SNe~Ib/c;][]{2011Li,2009Smarttetal}.
These are defined by a lack of hydrogen in their spectra, and are further differentiated based on the presence (SNe~Ib) or lack (SNe~Ic) of strong helium lines.
SNe~Ib/c, sometimes referred to as stripped-envelope supernovae, are thought to be the core-collapse explosions of massive stars that have lost their hydrogen (and in the case of SNe~Ic, helium) envelopes.
One particularly interesting subclass of stripped-envelope SNe are the broad-lined SNe~Ic (Ic-BL), which exhibit unusually broad absorption features and are associated with long-duration gamma-ray bursts \citep{1998Galama,2003Stanek}.
Further complicating the landscape of core-collapse taxonomy are the Type IIb supernovae that transition from resembling SNe~II at early times to exhibiting the narrow helium emission of SNe~Ib as they approach maximum light.
The nearby SN~1993J is the well-studied archetype of these transitional events \citep{1993Filippenko,1994Woosley,2004Maund,2020Stevance}.

Building a theoretical understanding of the rising light curves of core-collapse explosions has been an area of intense research for some time, in large part because constraining progenitor properties through archival observations is difficult (see the reviews by \citealt{2009Smartt} and \citealt{2015Smartt}).
\cite{2010Nakar} and \cite{2011Rabinak} explore semi-analytic approaches to this problem, assuming an idealized polytropic density profile for the stellar structure in order to make connections between the rising light curve and properties of the progenitor star.
Whether or not these idealized profiles actually occur at relevant depths within the star remains an open question \citep{2016Morozova}, and numerical models provide a means of studying these explosions with fewer restrictive physical assumptions.

Detailed simulations have been used to model observations of particularly well-studied events like SN~1987A \citep{1988Woosley,1993Utrobin,2019Utrobin}, SN~1993J \citep{1993Nomoto,2018Dessart}, SN~1999em \citep{2005Baklanov}, and SN~2011dh \citep{2012Bersten}.
Because these calculations are often tailored for the specific events, it is difficult to generalize their results.
To this end, many studies have tried to develop a more general understanding of how characteristics of the progenitor star impact the resultant supernova light curves.
Examples include \cite{2004Young}, \cite{2009Kasen}, \cite{2013Dessart}, \cite{2016Sukhbold}, and \cite{2020Curtis}, and we make extensive use of the model light curves from \cite{2016Morozova} here.
An excellent overview of our current understanding of core-collapse light curve physics is provided in Chapters 5 and 9 of \cite{2017BranchWheeler}.

Early-time observations of SNe~Ia have become increasingly accessible in recent years (e.g., \citealt{2016Shappee,2018Stritzinger,2019Yao,2019Fausnaugh}), but due to their lower intrinsic luminosities and faster rise times, the sample of core-collapse supernovae observed at early times remains small.
The first expected signature of these explosions occurs as the shock generated by the core-collapse approaches the surface of the progenitor, rapidly heating the photosphere to temperatures $\gtrsim 10^5$ K and producing a short-lived outburst of high-energy radiation with a duration comparable to the star's light crossing time  (e.g., \citealt{1978Klein,1992Ensman,2010Nakar}).
This shock breakout emission has been seen by the \textit{GALEX} satellite in the ultraviolet (UV) for a pair of SNe~II \citep{2008Schawinski,2015Gezari}, and in x-rays by the \textit{Swift} satellite for the Type Ib SN~2008D \citep{2008Soderberg,2009Modjaz}.
In principle this emission can also be seen in the optical and infrared (IR), but the weakness of the signature at these wavelengths coupled with its short duration ($\lesssim30$ min at peak) makes it difficult to observe for even the highest cadence ground-based surveys.

Observations from the \textit{Kepler} Space Telescope \citep{2010Haas} and its extended \textit{K2} mission were able to address both of these issues, as the mission provided high quality photometry at an unprecedented 30-minute cadence.
In fact, the \textit{Kepler} observations of the bright SN~Ia ASASSN-18bt provide the highest precision light curve of any supernova yet observed \citep{2019Shappee,2019Dimitriadis}.
Two \textit{Kepler} light curves of SNe~II have been published by \cite{2016Garnavich}, one of which shows a plausible detection of shock breakout emission.

In both the discovery of transiting exoplanets and the study of extragalactic transients, \textit{TESS} is the natural successor to \textit{Kepler}.
The high-cadence \textit{TESS} observations of the bright tidal disruption event ASASSN-19bt help to make it one of the best-studied optical TDEs to date \citep{2019Holoien}, and \cite{2019Vallely} and \cite{2019Fausnaugh} have used \textit{TESS} to study the early-time light curves of SNe~Ia.
In this paper we present observations of twenty bright core-collapse supernovae discovered over the course of the two year \textit{TESS} prime mission.
Although \textit{TESS} images are deep enough to be competitive with current ground-based discovery surveys, there is a considerable delay between when observations are taken and when they are downlinked from the spacecraft and subsequently made available to the public.
This delay is long enough that the mission is not an effective means of discovering bright long-lived transients.
As such, all of the events in our sample were discovered by other surveys.
It is critical that ground-based surveys quickly detect new transients in the \textit{TESS} fields in order to enable follow-up spectroscopic confirmation and classification as well as multi-band photometric monitoring.

For this reason, the All-Sky Automated Survey for Supernovae \citep[ASAS-SN;][]{2014ShappeeASASSN,2017Kochanek,2019Holoien} has been monitoring the \textit{TESS} observing fields with increased cadence since the mission began.
This monitoring program has proven very successful.
ASAS-SN discovered the unusual Type Ia supernova ASASSN-18tb, the first extragalactic transient to be studied with \textit{TESS} data \citep{2019Vallely}, the spectacular tidal disruption event ASASSN-19bt in the \textit{TESS} continuous viewing zone \cite{2019Holoien}, as well as a large fraction of the bright SNe~Ia from \textit{TESS} Sectors 1--6 studied by \cite{2019Fausnaugh}.
During the second year of the mission, the Zwicky Transient Facility \citep[ZTF;][]{2019Bellm,2019ZTFTESS} also adopted this practice, and as a result these two surveys are responsible for discovering the majority of the bright transients we study here.
Additional discoveries were made during the course of standard operations by the Asteroid Terrestrial-impact Last Alert System \citep[ATLAS;][2 SNe]{2018Tonry},  the  $D<40$ Mpc SN Survey  \citep[DLT40;][2 SNe]{2018Tartaglia}, the Mobile Astronomical System of TElescope Robots \citep[MASTER;][2 SNe]{2010Lipunov}, and from \textit{Gaia} alerts \citep[][1 SN]{2016Gaia}.

Table~\ref{tab:sample} summarizes the observational properties of the events in our sample, including their discovery dates, spectroscopic classifications, redshifts, and $V$-band Galactic extinctions \citep[$A_V$;][]{1998Schlegel,2011Schlafly}.
The bulk of the sample (12 of the total 20 events) is comprised of normal Type II SNe, and most of our subsequent analysis focuses on these events.
The rest of the sample is made up of three Type IIn SNe, and one each of Type IIb, Ib, Ic, Ic-BL, and Ibn SNe.
For inclusion in this study, we required that events have peak apparent \textit{TESS}-band magnitudes $\lesssim 18$ mag, and the majority of the light curves  we present reach peak magnitudes brighter than 17 mag.
The most impressive individual light curve in our sample is that of the Type IIn supernova DLT19c (SN~2019esa), which suffers from minimal \textit{TESS} observing gaps and attains a peak magnitude of $T_{mag}=13.83\pm0.01$.

In Section~\ref{sec:LIGER} we describe our \textit{TESS} image subtraction pipeline.
We discuss empirical fits to the observed core-collapse \textit{TESS} light curves in Section~\ref{sec:empirical}.
In Section~\ref{sec:modeling} we discuss semi-analytic models for the early-time light curves of the Type~II SNe and the resulting estimates of the progenitor radii and explosion energies.
We then compare our \textit{TESS} light curves to the numerically simulated light curves from \cite{2016Morozova} and investigate their use as calibrators for the semi-analytic models.
Section~\ref{sec:shockbreakout} discusses  shock breakout emission and its likely detection in the stacked light curves of the SNe~II in our sample.
Finally, we discuss our results in Section~\ref{sec:conclusions}.

\section{TESS Image Subtraction Light Curve Generation}
\label{sec:LIGER}

There are a number of challenges associated with the \textit{TESS} design that complicate the extraction of light curves, and considerable effort has been expended by the community to develop pipelines to address them.
The \textit{TESS} Science Processing Operations Center (SPOC) pipeline generates calibrated light curves and validation products that are used by the \textit{TESS} Science Office clearinghouse to identify promising transit candidates for additional follow-up.
This pipeline is focused primarily on identifying transiting exoplanet signatures in the 200,000 stars selected for monitoring at 2-minute cadence \citep{2016Jenkins,2018Stassun}.
The \textit{TESS} Asteroseismic Science Consortium (TASC) has produced its own pipeline, optimized for studying stellar oscillations, which computes light curves for all of the known stellar sources observed by \textit{TESS} \citep{2017Lund,2019Handberg}.
While these pipelines are designed to produce light curves for known sources, the open-source \textsc{eleanor} package offers a publicly available tool for producing light curves of any source observed by \textit{TESS} \citep{2019Feinstein}.

Ultimately, none of these pipelines are ideally suited for studying extragalactic transients with \textit{TESS}.
Because they generally are optimized for high-precision observations of short duration signals associated with bright ($T\lesssim15$ mag) stellar sources, their corrective procedures often weaken or entirely remove features that evolve on longer timescales.
This is a strength when searching for the fleeting flux dip during a planetary transit, but is clearly a problem when trying to study supernovae.

We have developed our own pipeline for the primary purpose of studying transients with \textit{TESS}.
Producing \textit{TESS} light curves using the image subtraction technique provides a natural means of addressing some of the otherwise troublesome characteristics of \textit{TESS} images.
Most notably, the large 21{\arcsec} pixels and undersampled point-spread function (PSF) can make source blending a serious challenge for data reductions based on conventional aperture or PSF photometry techniques.
This is not a problem for image subtraction, as light curves produced through this technique are sensitive only to changes in flux relative to the reference image.

We have applied many of the lessons learned over the course of operating the All-Sky Automated Survey for Supernovae \citep[ASAS-SN;][]{2014ShappeeASASSN,2017Kochanek} to develop an image subtraction pipeline optimized for use on the \textit{TESS} full-frame images (FFIs).
As in ASAS-SN data processing, the backbone of our \textit{TESS} light curve generator is the \textsc{ISIS} image subtraction package \citep{1998Alard,2000Alard}.
We work with 750-pixel wide ``postage stamps'' cut out of the full FFIs. These postage stamps are generally centered on the location of the target, but are off-centered by necessity when the target is less than 375 pixels from a detector edge.

The basic automated image subtraction procedure we use to process the \textit{TESS} FFIs broadly follows the standard \textsc{ISIS} package usage:
\begin{itemize}
  \item[1.] Produce 750-pixel postage stamps from each of the FFIs
  \item[2.] Interpolate all images to the same astrometric solution
  \item[3.] Select 100 high quality images for building the reference image
  \item[4.] Scale and subtract this reference image from all postage stamps
  \item[5.] Perform PSF photometry on the subtracted images
\end{itemize}

Images used to build the reference image must pass two tests.
First, the associated FFI must not have any mission-provided data quality flags.
This ensures that we avoid images compromised by momentum dumps and other issues described in the \textit{TESS} Science Data Product Description Document\footnote{\url{https://archive.stsci.edu/missions/tess/doc/EXP-TESS-ARC-ICD-TM-0014.pdf}}.
Second, the image must have PSF widths and sky background levels at or below the median values measured from all of that sector's images.
This second test is important for excluding images with moderately elevated background counts due to scattered light from the Earth and Moon, as these FFIs are not always flagged by the mission.
While they are not used to construct the reference image, these images can still be used in the light curves.

In general, this basic approach is fairly reliable.
We used it to produce the first published \textit{TESS} supernova light curve \citep{2019Vallely}, observe the first promising repeating partial tidal disruption event candidate ASASSN-14ko \citep{2020Payne}, and study a sample of Southern Hemipshere $\delta$ Scuti stars \citep{2019JayasingheDScuti}.
However, it occasionally yields inconsistent results or artificial trends.
One problem is the undersampled PSF of the \textit{TESS} images.
This makes it useful in some cases to convolve the images with a Gaussian prior to performing the image subtraction.
Since this Gaussian smearing is flux-conserving, the only intrinsic drawback of the technique is that it slightly lowers the signal-to-noise ratio of the resulting light curve.

Light curves produced through this Gaussian smearing method have been used to study the tidal disruption event ASASSN-19bt \citep{2019Holoien}, characterize the short-term variability of ESO-H$\alpha$ 99's EXor outburst \citep{2019Hodapp}, and detect an 8.66-hour periodicity in the light curve of V1047 Cen \citep{2019Aydi}.
One needs to be careful when using this technique, however.
For targets located near other strongly variable sources, for example, applying the additional smearing will exacerbate signal pollution from the neighboring source.
Similar issues also arise if the target is close to a bright source, as the subtraction artifacts associated with the bright source can significantly degrade the target light curve after the Gaussian smearing is applied.
Given the large pixel scale of \textit{TESS}, these issues are not uncommon, so care must be taken on a sector-by-sector basis to gauge the utility of this correction.

While the Gaussian smearing technique is useful for mitigating issues with the undersampled PSF, the most significant problem with \textit{TESS} observations have proven to be background artifacts produced by scattered light from the Earth and Moon (See, e.g., Section 7.32 of the \textit{TESS} Instrument Handbook\footnote{\url{https://archive.stsci.edu/files/live/sites/mast/files/home/missions-and-data/active-missions/tess/_documents/TESS_Instrument_Handbook_v0.1.pdf}}) and the ``strap'' artifacts.
These strap artifacts are produced by reflective metal straps located at the base of the silicon depletion region of the \textit{TESS} detectors, and are most strongly present during epochs of significant scattered light (see Section 6.6.1 of the TESS Instrument Handbook for more details).
The basic \textsc{ISIS} image subtraction procedure incorporates a 2-dimensional polynomial sky model to remove differential background variations, but this sometimes proves insufficient for the complex scattered light structures present in \textit{TESS} images.
After a talk by Marco Montalto at TESS Science Conference I, we were inspired to investigate the utility of including an additional round of background filtering in our image processing sequence.
Ultimately, this led us to develop the following procedure, which has proven very effective at addressing these two issues.

We implement a procedure that utilizes a one-dimensional median filter, applied once along each axis of the CCD, so that it captures background structure in both the vertical and horizontal directions of the detector.
This dual median filtering is a simple, but effective, image processing technique that is applied to each postage stamp following the initial \textsc{ISIS} image subtraction.
Our technique is described by the following steps:
\begin{itemize}
  \item[1.] Obtain the vertical (i.e., parallel to the straps) background template by measuring the median value in 30 pixel windows slid along each column of the postage stamp
  \item[2.] Subtract this vertical background template from the postage stamp, producing an image that has been median filtered along its columns
  \item[3.] Obtain the horizontal background template by measuring the median value in 30 pixel windows slid along each row of the 1-D median filtered image produced in Step 2.
  \item[4.] Subtract this horizontal background template from the 1-D median filtered image to produce an image that has been median filtered along both axes
  \item[5.] Perform PSF photometry on the median filtered image produced in Step 4.
\end{itemize}

\begin{figure*}
\centering
\rotatebox[origin=c]{90}{Full Postage Stamp}
\begin{subfigure}{.32\textwidth}
 \centering
 \includegraphics[width=\textwidth]{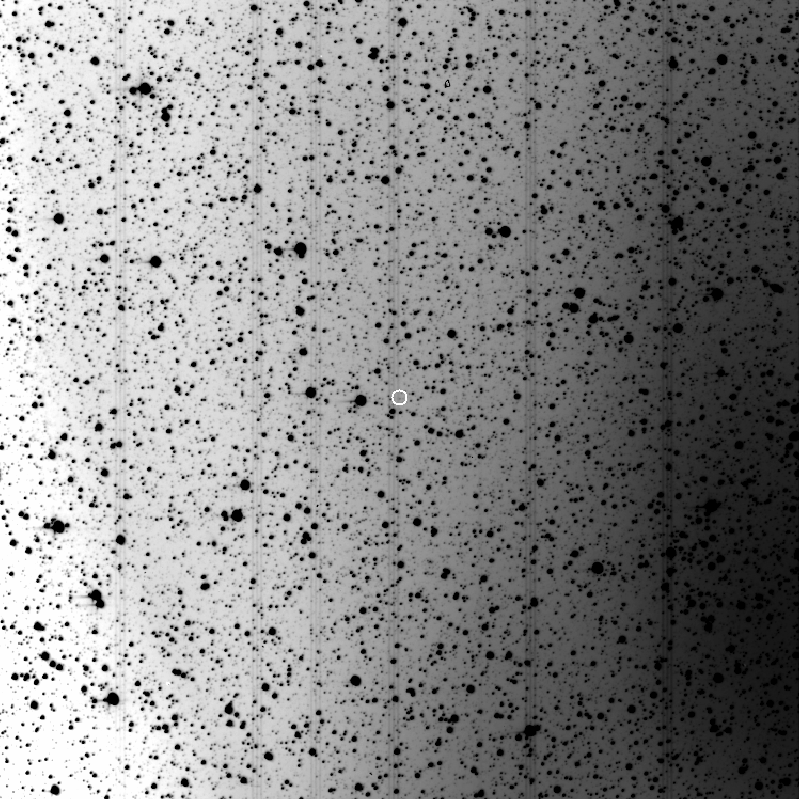}
\end{subfigure}
\begin{subfigure}{.32\textwidth}
 \centering
 \includegraphics[width=\textwidth]{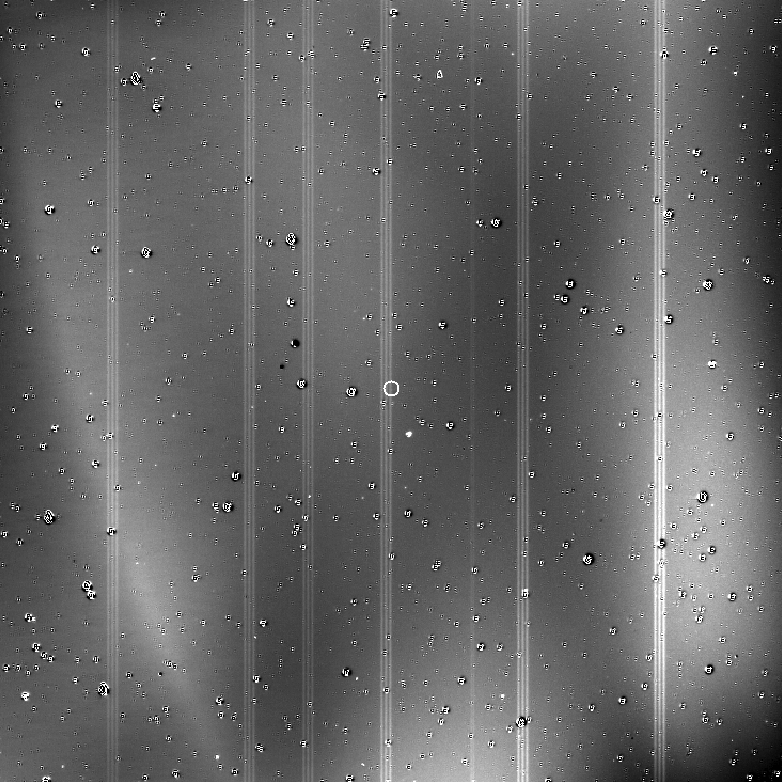}
\end{subfigure}
\begin{subfigure}{.32\textwidth}
 \centering
 \includegraphics[width=\textwidth]{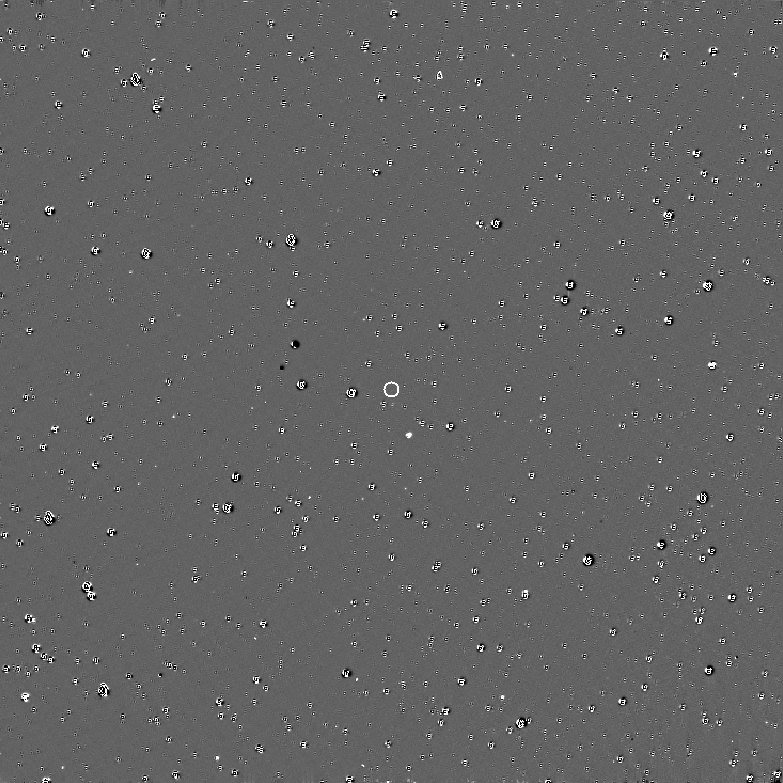}
\end{subfigure}
\rotatebox[origin=c]{90}{Local Target Region}
\begin{subfigure}{.32\textwidth}
 \includegraphics[width=\textwidth,right]{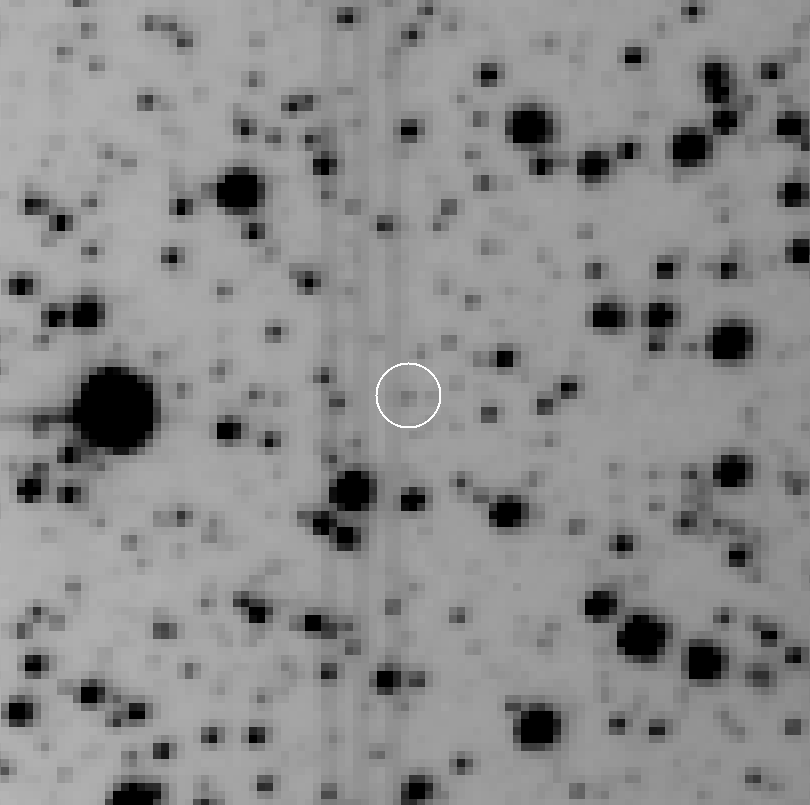}
 \caption{Unprocessed \textit{TESS} FFI}
\end{subfigure}
\begin{subfigure}{.32\textwidth}
 \includegraphics[width=\textwidth,right]{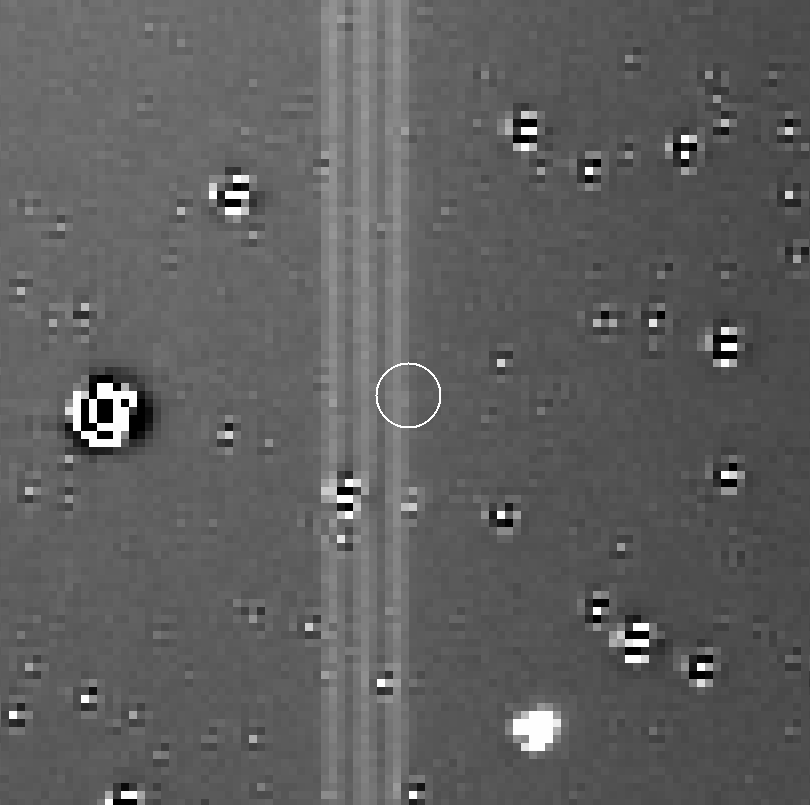}
 \caption{Initial Image Subtraction}
\end{subfigure}
\begin{subfigure}{.32\textwidth}
 \includegraphics[width=\textwidth,right]{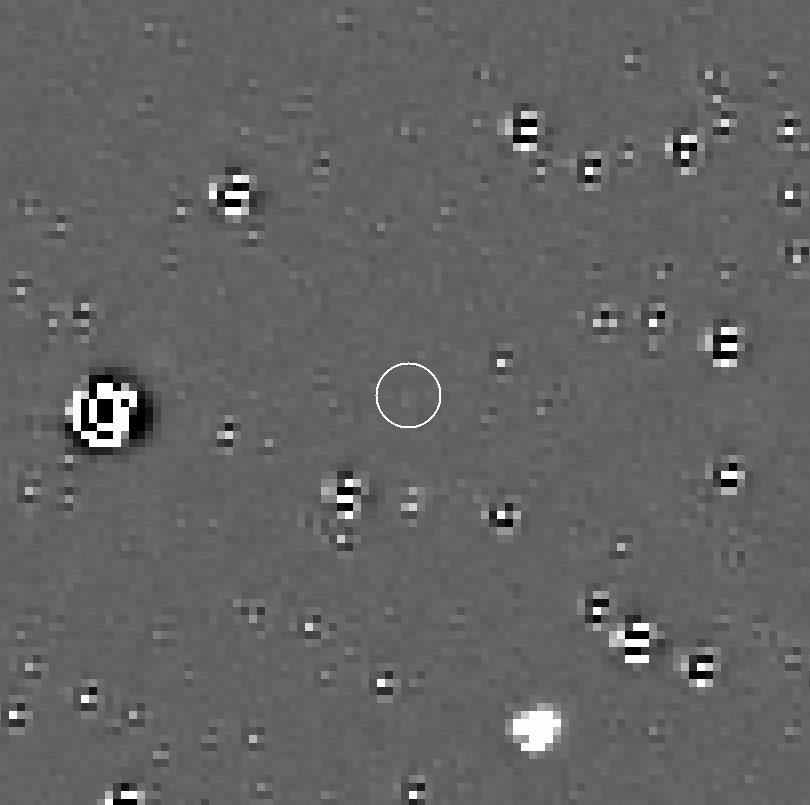}
 \caption{After Median Filtering}
\end{subfigure}
\caption{The image processing sequence for the \textit{TESS} observation of ASASSN-18qk obtained on JD 2458341.51 by Camera~1/CCD~3. Column (a) shows the FFI as available for download from MAST, column (b) shows the subtracted image produced using the basic \textsc{ISIS} image subtraction, and column (c) shows the image produced after applying the median filtering technique described in Section~\ref{sec:LIGER}. The images in columns (b) and (c) are shown using the same scale, which has a dynamic range about 25 times smaller than that used in column (a). The top row shows the full 750-pixel wide postage stamps, while the bottom row is now zoomed in on a $100\times100$ pixel region centered on the location of ASASSN-18qk (circled in white). While the basic image subtraction process is able to effectively remove the large scale gradient visible in the FFI by fitting a low order polynomial to the differential background, it is clear that some of this background structure and the strap artifacts (the vertical lines) remain in the initial image subtraction. Both of these contaminants are removed after applying the median filtering technique.}
\label{fig:imagesequence}
\end{figure*}

We favor this double one-dimensional filter approach rather than a single application of a two-dimensional filter because it provides an excellent means of removing CCD strap artifacts from the images.
The initial column filtering allows us to remove the strap artifacts rather than blending them into the surrounding pixels, as occurs when a two-dimensional filter is used.
Because the straps lie precisely along detector columns, the one dimensional filter applied in the vertical direction is able to effectively remove them.
An example of this image processing sequence is shown in Figure~\ref{fig:imagesequence}.
Although developed independently, we note that this technique is similar to that utilized by \cite{2020Pal} for studying asteroids and other solar system objects with \textit{TESS}.

\begin{figure*}
\centering
\includegraphics[width=0.32\textwidth]{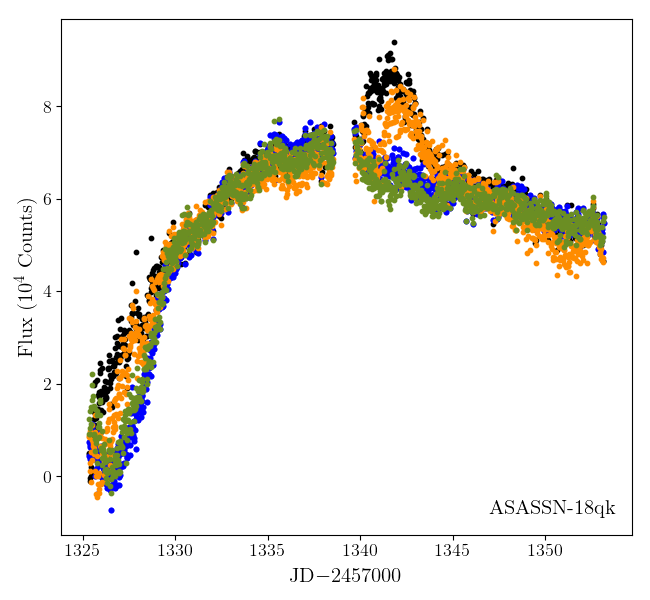}
\includegraphics[width=0.32\textwidth]{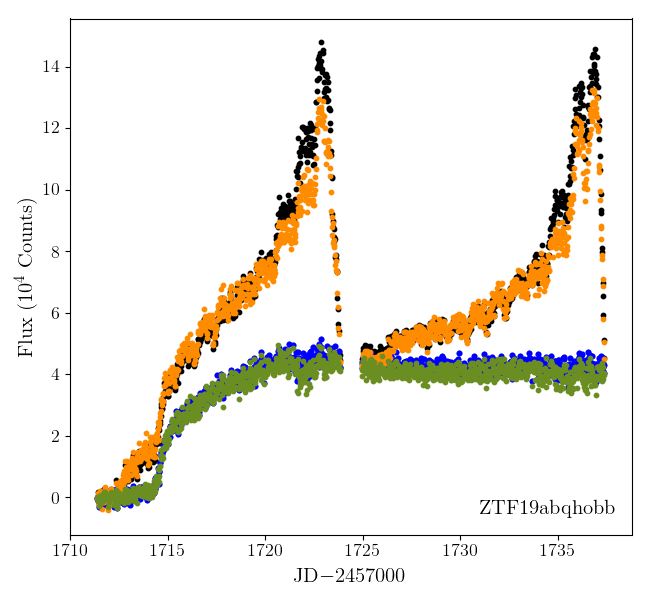}
\includegraphics[width=0.32\textwidth]{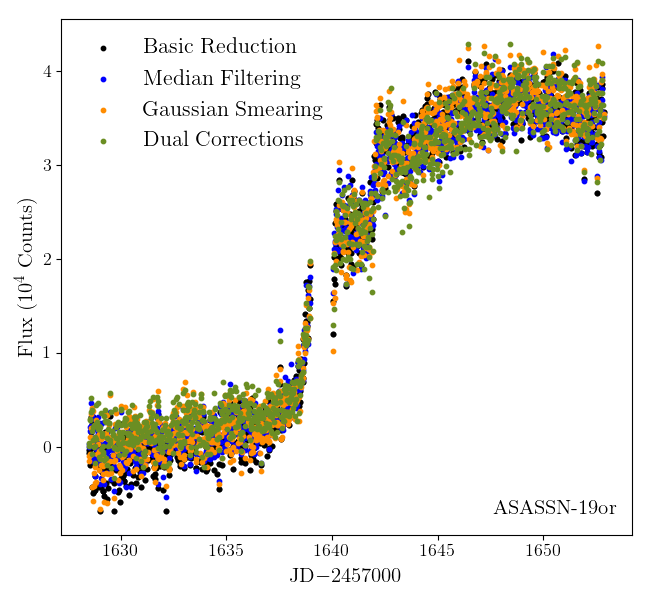}
\caption{The four light curves produced by our automated image subtraction pipeline for the \textit{TESS} observations of the supernovae ASASSN-18qk (left), ZTF19abqhobb (middle), and ASASSN-19or (right). In the cases of ASASSN-18qk and, particularly, ZTF19abqhobb, it is clear that incorporating median filtering in the processing sequence greatly improves the light curve quality. In some cases, like ASASSN-19or, however, there is no clear benefit to the additional processing steps.}
\label{fig:LIGEReffectiveness}
\end{figure*}

We use 30 pixels as the default width for the one-dimensional median filter, although this value can be easily increased as needed to accommodate brighter targets.
In \textit{TESS} images, 90\% of the total flux of a point-source is located within a $2\times2$ pixel area \citep{2015Ricker}.
Since the width of the median filter is considerably larger than 2 pixels and we are interested in relatively faint sources, the filtering should not --- and does not appear to --- remove appreciable amounts of source flux from the image.
Whenever possible, the validity of this assumption is confirmed for each target by comparing the light curve measured after median filtering with that of the base-level reduction (see Figure~\ref{fig:LIGEReffectiveness}).
For cases where it is evident that source flux was removed during the median filtering, the procedure is repeated using a median filter of larger width.

For each target, our automated image subtraction pipeline produces four light curves.
The first is a basic \textsc{ISIS} reduction with neither of the \textit{TESS}-specifc corrective techniques applied.
The second and third are produced by either applying the Gaussian smearing or median filtering corrections separately.
The final light curve is produced by applying both techniques simultaneously.
All four light curves are subsequently inspected manually, and the most appropriate reduction is selected for further analysis.
Due to the reduced signal-to-noise inherent to the technique, the light curves that utilize Gaussian smearing are only considered further if they appear necessary.
Our preferred ordering is: 1) basic \textsc{ISIS} reduction, 2) median filtering only, 3) Gaussian smearing only, 4) both Gaussian smearing and median filtering.

Figure~\ref{fig:LIGEReffectiveness} illustrates how important these corrective techniques can be, using the light curves of ASASSN-18qk, ZTF19abqhobb, and ASASSN-19or.
For all three events the basic \textsc{ISIS} reduction is shown in black, the Gaussian smearing reduction is shown in orange, the median filtering reduction is shown in blue, and the dual correction reduction is shown in green.
The importance of the median filtering is particularly evident in ZTF19abqhobb, as the two reductions that do not include this correction are completely dominated by background artifacts.
It is not clear, however, that the additional Gaussian smearing utilized in the dual correction reduction mitigates any artifacts not already removed by the median filtering.
Thus, for this target we select the median filter only reduction for subsequent analysis.

The differential light curves we produce measure changes in flux relative to the reference image.
For transients like the supernovae we study here, one would prefer to build a reference image exclusively using observations obtained long before the transient's rise.
This ensures that no flux from the transient is present in the reference image.
However, the relatively short 27 day monitoring window set by the \textit{TESS} observing strategy means that this is not always possible.
We address this concern by measuring the median value of the pre-rise flux measurements for each light curve and subtracting this flux value from all epochs.
Introducing this flux offset shifts the entire light curve, ensuring it has a baseline of zero flux prior to the rise of the transient.

For transients observed by \textit{TESS} over multiple sectors, it is in principle possible to generate a single reference image and then rotate it for use with each new pointing.
In practice, however, the large pixel scale of the \textit{TESS} observations makes this particularly difficult and introduces a large source of uncertainty.
As in our prior work (e.g.,\citealt{2019Vallely,2019Holoien,2020Tucker}), we instead choose to construct independent reference images for each sector.
When doing so, the flux offset of the first sector is determined from the pre-rise zero point described above.
For subsequent sectors, the flux offset is chosen such that linear extrapolations from the last $\sim2$ days of the earlier sector and the first $\sim2$ days of the later sector match.

In addition to the scattered light and strap artifacts described earlier, \textit{TESS} images can be compromised by a number of issues, including the spacecraft's frequent momentum dumps, on board instrument anomalies, or ghost artifacts produced by nearby bright sources.
Most of these issues are fairly well understood and are discussed in the \textit{TESS} Instrument Handbook and \textit{TESS} Data Release Notes\footnote{\url{https://archive.stsci.edu/tess/tess_drn.html}} documentation.
While many of these conditions are noted by the mission provided quality flags that accompany the publicly available FFIs, it is often the case that non-flagged images are still compromised.
For instance, the ASASSN-18qk FFI shown in Figure~\ref{fig:imagesequence} is not flagged by the mission.

The final step of our data reduction process is a manual vetting of the light curves to identify and remove any remaining compromised epochs.
This is a somewhat subjective process, in which we examine light curves for epochs dominated by uncorrected systematics and remove them.
In practice, these compromised epochs almost always coincide with either momentum dumps or issues with the observations documented in the Data Release Notes for the sector in question.
We do not use photometry from these compromised epochs in our analysis, but for completeness they are included, but flagged, in the \textit{TESS} light curves that we provide in the online supplementary material.

\begin{table*}
\caption{The Twenty Core-Collapse Supernovae in Our Sample.}
\label{tab:sample}
\begin{tabular}{cllllc}
\hline\hline
Supernova & TNS ID & Discovery Date & Spectral Classification & Redshift & $A_V$ (mag) \\
\hline\hline
ASASSN-18qk & 2018emt & 2018-08-01 \citep{18qkDisc} & II \citep{18qkClass} & 0.02395  & 0.068 \\
ASASSN-18qv & 2018eph & 2018-08-04 \citep{18qvDisc} & II \citep{18qvClass} & 0.03  & 0.066 \\
ZTF18abzscns & 2018gxi & 2018-09-28 \citep{ZTF18abzscnsDisc} & II \citep{ZTF18abzscnsClass} & 0.057 & 0.055 \\
\makecell{MASTEROT\\J065447.10$-$593010.8} & 2018jmt & 2018-12-08 \citep{18jmtDisc} & Ibn \citep{18jmtClass} & 0.036 & 0.327 \\ 
DLT18ar & 2018lab & 2018-12-29 \citep{18labDisc} & II \citep{18labClass} & 0.0092 & 0.238 \\
ATLAS19dzi & 2019bwb &  2019-03-17 \citep{19dziDisc} & IIn \citep{19dziClass} & 0.02019  & 0.536 \\ 
ASASSN-19jy & 2019dke &  2019-04-11 \citep{19jyDisc} & II \citep{19jyClass} & 0.01064  & 0.227 \\ 
ATLAS19giz & 2019dhz & 2019-04-04 \citep{19gizDisc} & II \citep{19gizClass} & 0.034 & 0.214 \\
DLT19c & 2019esa &  2019-05-06 \citep{19esaDisc} & IIn \citep{19esaClass} & 0.00589  & 0.496 \\ 
\makecell{MASTEROT\\J135130.87$-$525534.4} & 2019fcc & 2019-05-12 \citep{19fccDisc} & II \citep{19fccClass} & 0.0126 & 1.079 \\
ASASSN-19or & 2019hcn &  2019-06-08 \citep{19orDisc} & II \citep{19orClass} & 0.018  & 0.309 \\ 
Gaia19dcu & 2019lqo & 2019-07-21 \citep{19dcuDisc} & II \citep{19dcuClass} & 0.0103 & 0.046 \\
ZTF19abqhobb & 2019nvm &  2019-08-19 \citep{19nvmDisc} & II \citep{19nvmClass} & 0.01815  & 0.082 \\ 
ZTF19abvdgqo & 2019pfb & 2019-09-01 \citep{ZTF19abvdgqoDisc} & Ib \citep{ZTF19abvdgqoClass} & 0.03683 & 0.223 \\
ASASSN-19acc & 2019vxm & 2019-12-01 \citep{19accDisc} & IIn \citep{19accClass} & 0.019  & 0.282 \\
ZTF20aagnbes & 2020aem & 2020-01-18 \citep{ZTF20aagnbesDisc} & II \citep{ZTF20aagnbesClass} & 0.02218 & 0.116 \\
ZTF20aahbamv & 2020amv & 2020-01-23 \citep{20cowDisc} & II \citep{20cowClass} & 0.0452 & 0.100 \\
ZTF20aatzhhl & 2020fqv & 2020-03-31 \citep{20fqvDisc} & IIb \citep{20fqvClass} & 0.00752 & 0.089 \\
ASASSN-20dn & 2020euy & 2020-03-23 \citep{20dnDisc} & Ic \citep{20dnClass} & 0.06231 & 0.063 \\
ZTF20abbplei & 2020lao & 2020-05-25 \citep{ZTF20abbpleiDisc} & Ic-BL \citep{ZTF20abbpleiClass} & 0.03116 & 0.138  \\
\hline\hline
\end{tabular} \\
\end{table*}

\begin{figure*}
    \centering
    \includegraphics[width=0.96\textwidth]{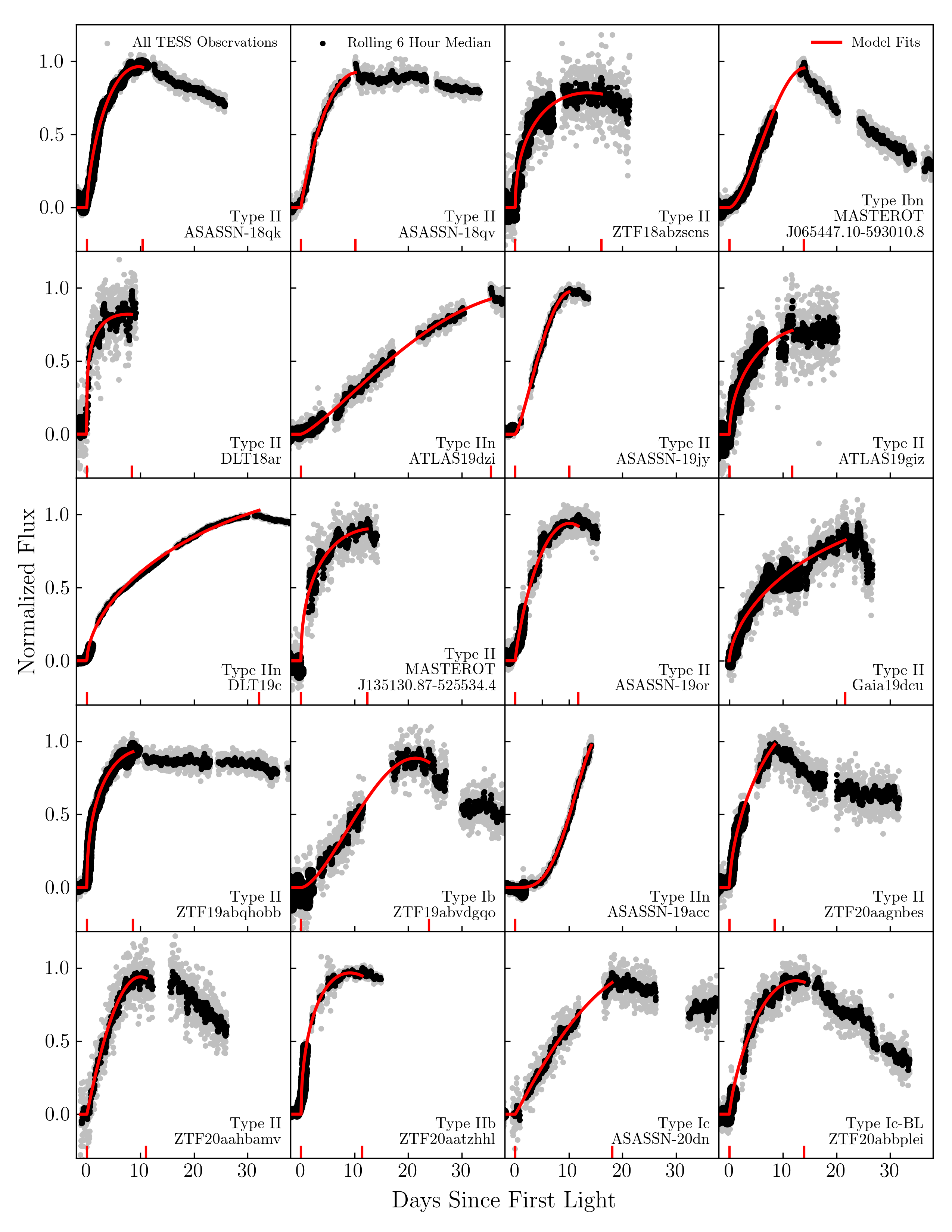}
    \caption{\textit{TESS} light curves of the twenty core-collapse events in our sample. For each event, all of the \textit{TESS} observations are shown in gray, with their rolling 6 hour median shown in black, and the best-fit model (Equation~\ref{eq:CurvedPL}) for the rising light curve is shown in red. The two red tick marks at the bottom of each panel indicate the times of first and maximum light for each light curve. All of these light curves are included in machine-readable format in the online supplementary material.}
    \label{fig:AllSNe}
\end{figure*}

\begin{table*}
\caption{\textit{TESS} Light Curve Parameters.}
\label{tab:params}
\begin{tabular}{cccccccc}
\hline\hline
Supernova & Peak Mag. & Peak Mag. & Rise Time & $t_1$ & $\alpha$ & $a_1$ & $a_2\cdot100$ \\
 & (Apparent) & (Absolute) & (Days) & (TJD) & & & (Days$^{-1}$) \\
\hline\hline
ASASSN-18qk & 16.19$\pm$0.04 & $-$18.85$\pm$0.10 & 10.2$\pm$0.21 & 1327.40$\pm$0.02 & 1.62$\pm$0.25 & 0.68$\pm$0.02 & $-$4.27$\pm$0.14 \\
ASASSN-18qv & 16.31$\pm$0.04 & $-$19.22$\pm$0.08 & \hspace{4.5pt}9.9$\pm$0.21 & 1329.73$\pm$0.04 & 2.39$\pm$0.46 & 0.88$\pm$0.03 & $-$4.61$\pm$0.20 \\
ZTF18abzscns & 18.05$\pm$0.15 & $-$18.92$\pm$0.15 & 15.2$\pm$0.21 & 1387.45$\pm$0.05 & 0.73$\pm$0.33 & 0.45$\pm$0.03 & $-$2.41$\pm$0.28 \\
\makecell{MASTEROT\\J065447.10$-$593010.8} & 16.86$\pm$0.06 & $-$19.22$\pm$0.09 & 13.4$\pm$0.32 & 1455.51$\pm$0.24 & 2.45$\pm$0.38 & 1.68$\pm$0.11 & $-$4.60$\pm$0.18 \\
DLT18ar & 17.55$\pm$0.16 & $-$15.48$\pm$0.29 & \hspace{4.5pt}8.3$\pm$0.21 & 1480.90$\pm$0.01 & 1.44$\pm$1.21 & 0.23$\pm$0.03 & $-$2.43$\pm$0.76 \\
ATLAS19dzi & 16.33$\pm$0.04 & $-$18.56$\pm$0.12 & 34.7$\pm$0.34 & 1551.30$\pm$0.27 & 1.39$\pm$0.09 & 1.28$\pm$0.04 & $-$1.32$\pm$0.08 \\
ASASSN-19jy & 15.80$\pm$0.02 & $-$17.61$\pm$0.20 & 10.0$\pm$0.24 & 1581.98$\pm$0.12 & 2.55$\pm$0.77 & 1.34$\pm$0.05 & $-$5.58$\pm$0.13 \\
ATLAS19giz & 18.06$\pm$0.20 & $-$17.83$\pm$0.21 & 11.3$\pm$0.23 & 1575.48$\pm$0.09 & 0.46$\pm$0.12 & 0.53$\pm$0.08 & $-$2.16$\pm$0.97 \\
\makecell{MASTEROT\\J135130.87$-$525534.4} & 16.99$\pm$0.09 & $-$17.26$\pm$0.19 & 12.2$\pm$0.21 & 1609.65$\pm$0.01 & 0.10$\pm$0.04 & 0.44$\pm$0.03 & $-$2.23$\pm$0.39 \\
DLT19c & 13.83$\pm$0.01 & $-$18.40$\pm$0.37 & 32.0$\pm$0.21 & 1608.94$\pm$0.01 & 0.63$\pm$0.01 & 0.57$\pm$0.01 & $-$0.53$\pm$0.02 \\
ASASSN-19or & 16.86$\pm$0.05 & $-$17.67$\pm$0.13 & 11.5$\pm$0.22 & 1637.49$\pm$0.06 & 1.59$\pm$0.29 & 0.83$\pm$0.03 & $-$4.57$\pm$0.16 \\
Gaia19dcu & $<$16.51 & $<-$16.65 & $>$21.4 & 1683.45$\pm$0.10 & 0.64$\pm$0.08 & 0.59$\pm$0.07 & $-$0.76$\pm$0.32 \\
ZTF19abqhobb & 16.64$\pm$0.03 & $-$17.79$\pm$0.13 & \hspace{4.5pt}8.4$\pm$0.21 & 1714.34$\pm$0.01 & 1.09$\pm$0.13 & 0.44$\pm$0.01 & $-$2.91$\pm$0.26 \\
ZTF19abvdgqo & 17.11$\pm$0.10 & $-$18.96$\pm$0.12 & 23.0$\pm$0.55 & 1721.88$\pm$0.51 & 0.75$\pm$0.09 & 1.65$\pm$0.10 & $-$3.02$\pm$0.07 \\
ASASSN-19acc & $<$15.63 & $<-$19.01 & $>$13.9 & 1800.79$\pm$0.48 & 1.51$\pm$0.03 & 3.52$\pm$0.23 & $-$5.05$\pm$0.07 \\
ZTF20aagnbes & 17.27$\pm$0.07 & $-$17.62$\pm$0.12 & \hspace{4.5pt}8.3$\pm$0.21 & 1866.02$\pm$0.03 & 2.17$\pm$0.81 & 0.61$\pm$0.04 & $-$1.78$\pm$0.84 \\
ZTF20aahbamv & 17.60$\pm$0.13 & $-$18.87$\pm$0.14 & 10.6$\pm$0.26 & 1871.67$\pm$0.16 & 0.98$\pm$0.24 & 1.02$\pm$0.07 & $-$5.25$\pm$0.30 \\
ZTF20aatzhhl & 14.77$\pm$0.02 & $-$17.74$\pm$0.29 & 11.3$\pm$0.21 & 1939.98$\pm$0.01 & 2.53$\pm$0.56 & 0.49$\pm$0.01 & $-$3.62$\pm$0.11 \\
ASASSN-20dn & 17.26$\pm$0.06 & $-$19.91$\pm$0.07 & 17.0$\pm$0.30 & 1914.61$\pm$0.21 & 0.99$\pm$0.10 & 1.05$\pm$0.07 & $-$2.05$\pm$0.28 \\
ZTF20abbplei & 16.77$\pm$0.07 & $-$18.89$\pm$0.10 & 13.5$\pm$0.22 & 1994.63$\pm$0.06 & 0.76$\pm$0.11 & 0.79$\pm$0.03 & $-$3.66$\pm$0.16 \\
\hline\hline
\end{tabular} \\
\begin{flushleft}
Absolute magnitudes reported here account for Galactic extinction and include a 300 km/s uncertainty due to peculiar velocities, except for the three events at $z<0.01$ for which we have used the \cite{2007Theureau} Tully-Fisher relation distance estimates. TJD is a standard time for \textit{TESS} data, defined as JD $-$ 2,457,000.0 days. We exclude Gaia19dcu and ASASSN-19acc from our analysis because their rising \textit{TESS} light curves are incomplete.
\end{flushleft}
\end{table*}

\section{Empirical Analysis of  Light Curves}
\label{sec:empirical}

\begin{figure*}
    \centering
    \includegraphics[width=\textwidth]{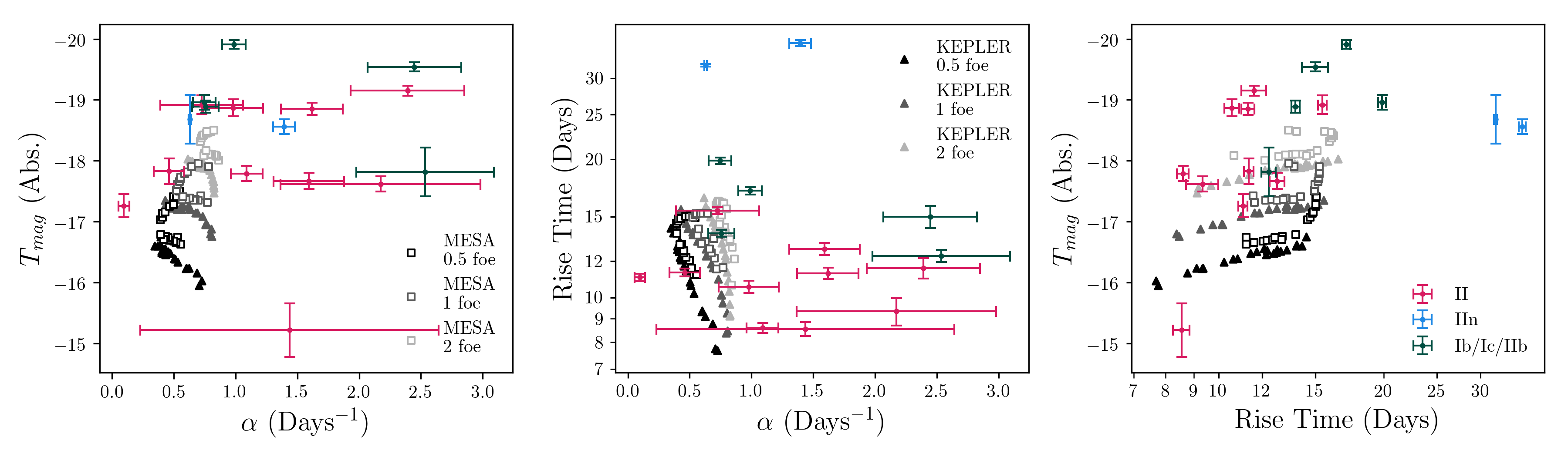}
    \caption{Single-component power-law fit parameters compared to rise time and \textit{TESS}-band absolute magnitude for the core-collapse events in our sample as well as a suite of models computed by \protect\cite{2016Morozova} from which we calculate synthetic \textit{TESS} observations. The colored points indicate observed data, with SNe~II in dark red, SNe~IIn in light blue, and SNe~Ib/c and IIb in dark green. The shaded points indicate model light curves. Progenitors simulated using \textsc{MESA} are shown as open squares, and those simulated using \textsc{KEPLER} are shown as solid triangles. Explosion energy is indicated by shading, with lighter shades corresponding to higher explosion energy.}
    \label{fig:SingleCompParameters}
\end{figure*}

In the literature, light curve rises are commonly characterized using a single-component power-law, $f(t) \propto t^\alpha$ (See, e.g., \citealt{Olling2015}).
While we prefer a slightly different treatment (described below), to facilitate comparisons with existing studies we first fit the light curves as
\begin{equation}
\label{eq:SinglePL}
  f(t) = \frac{h}{(1+z)^2} \bigg(\frac{t-t_1}{1+z}\bigg)^{\alpha} + f_0.
\end{equation}
for $t>t_1$ and as $f(t)=f_0$ for $t<t_1$.  Here $f_0$ is any residual background flux and $t_1$ is the beginning of the model rise.
The factors of $1+z$ are introduced to account for redshift time-dilation, although this is a relatively small effect for this sample since all of the events are found at low redshift ($z<0.063$).
Following \cite{Olling2015} and \cite{2019Fausnaugh}, we only fit the light curves up to 40\% of their peak flux.
The resulting fit parameters are shown in Figure~\ref{fig:SingleCompParameters}, and best-fit $\alpha$ values are presented in Table~\ref{tab:params}.

A single power law must eventually diverge from the actual light curve, forcing a somewhat arbitrary choice of a time to truncate the single power-law fits.
We can minimize this problem by using curved power-law fits of the form
\begin{equation}
\label{eq:CurvedPL}
f(t) = \frac{h}{(1+z)^2} \bigg(\frac{t-t_1}{1+z}\bigg)^{a_1\big(1+a_2\cdot(t-t_1)/(1+z)\big)} + f_0,
\end{equation}
up to (near) the light curve peak.  The $t_1$ and $f_0$ parameters are the same, and at early times
these fits become a $f~\propto~t^{a_1}$ power-law.
The $a_2$ term allows the model to follow the curvature of the light curve towards peak and so minimizes biases in estimates of the early time power-law exponent $a_1$.
Mathematically, $a_2$ is related to the rise time, $t_{rise}=t_{peak}-t_1$, between the start of the rise at time $t_1$ and the time of the peak $t_{peak}$ by
\begin{equation}  
    a_2^{-1} = - t_{rise}\left[ 1 +\ln\left(t_{rise}\right)\right].
\end{equation}
Fitting $a_2$ or some other variant of $t_{rise}^{-1}$ has better error characteristics than trying to fit a parameter like $a_2^{-1}$ because some fits allow $a_2$ values consistent with zero when the TESS light curve does not include the peak.  For our estimates of $t_{rise}$, we actually estimate $t_{peak}$ directly from the light curves as the time of peak flux in a rolling 6 hour binned light curve rather than from the
$a_2$ values because we still only fit the curved power laws to near the time of peak and not over the peak -- this would require a model with more parameters.  Essentially, using the curved power law parameter $a_2$ to determine $t_{rise}$ suffers from the same sorts of biases as using the single power law does for estimates of $a$.  For most fits, however, the estimates of
$t_{rise}$ from $a_2$ are in rough agreement with our more direct estimates.

\begin{figure*}
    \centering
    \includegraphics[width=\textwidth]{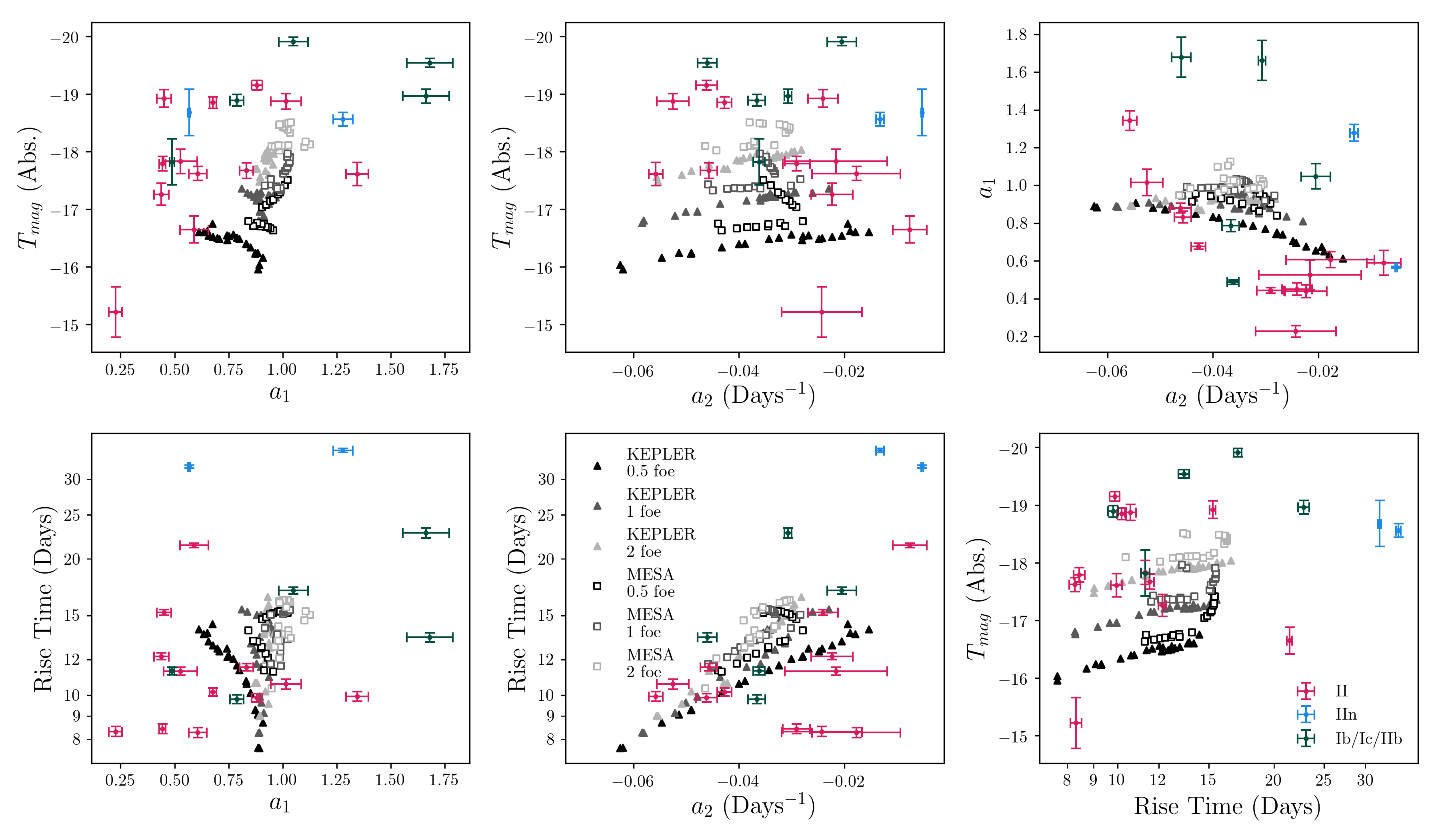}
    \caption{Curved power-law fit parameters compared to rise time and \textit{TESS}-band absolute magnitude for the core-collapse events in our sample as well as a suite of models computed by \protect\cite{2016Morozova} from which we calculate synthetic \textit{TESS} observations. Marker colors, shapes, and shades have the same meaning as in Figure~\ref{fig:SingleCompParameters}.}
    \label{fig:PLParameters}
\end{figure*}

The curved power-law fits are shown by the red curves in Figure~\ref{fig:AllSNe}, and Figure~\ref{fig:PLParameters} shows the peak absolute magnitude and rise time measured by \textit{TESS} for each event as a function of the two curved power-law indexes, $a_1$ and $a_2$.
Different colors are used to differentiate between spectroscopic subtypes, although the only clear difference among them is that the CSM-interacting SNe~IIn (shown in blue) exhibit substantially longer rise times.
We exclude two events from this comparison: ASASSN-19acc because the \textit{TESS} observations do not span the full rise to peak brightness, and Gaia19dcu because it was already rising at the start of its \textit{TESS} observations.
We still obtain fits for these two events, and for completeness their parameters are reported along with those of the rest of the sample in Table~\ref{tab:params}.
For comparison, Figure~\ref{fig:PLParameters} also includes the results from fitting the same curved power-law to the numerical model light curves from \cite{2016Morozova}.
These models are discussed in detail in Section~\ref{subsec:numerical}.

\begin{figure}
    \centering
    \includegraphics[width=\columnwidth]{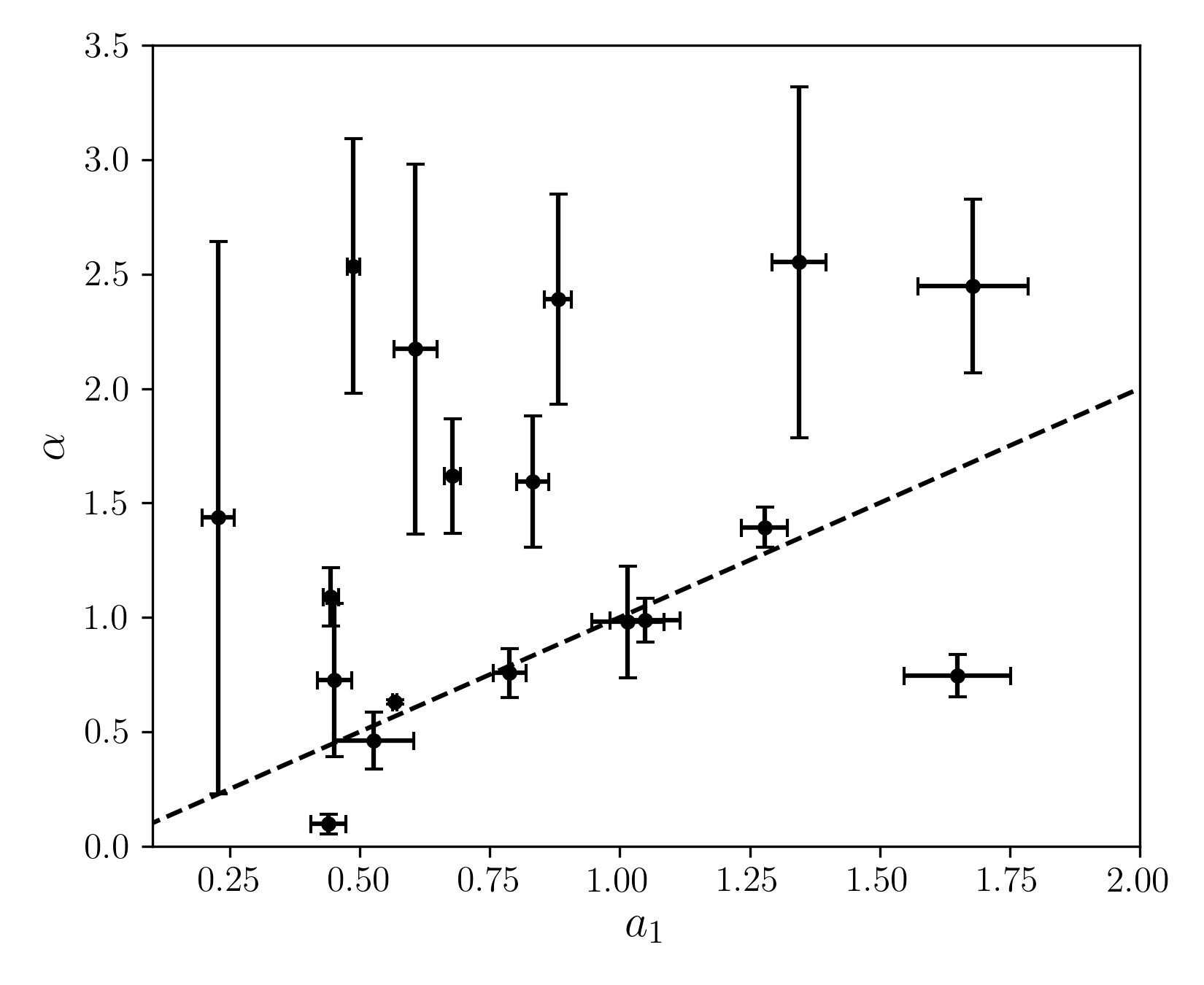}
    \caption{Comparing the initial rise indices from the single-component ($\alpha$) and curved power-law ($a_1$) forms. The dashed line indicates $a_1$=$\alpha$. For well-constrained values of $\alpha$ the two fits agree quite well, but for poorly constrained values the single-component fits are consistently larger than those from the curved power-law fits.}
    \label{fig:RiseIndices}
\end{figure}

The single-component power-law's $\alpha$ parameter and the curved power-law's $a_1$ parameter should be closely related, and Figure~\ref{fig:RiseIndices} compares their best-fit values for the well-observed events in our sample.
For light curves where $\alpha$ is well-constrained, the best-fit $\alpha$ values agree very well with the best-fit $a_2$ values.
In cases where $\alpha$ is not well-constrained, however, the best-fit $\alpha$ values are consistently and systematically larger than the best-fit $a_1$ values.
This suggests that, in addition to the avoidance of an arbitrary end point, curved power-law fits may provide a more robust means of characterizing early-time light curves than the single-component power-law (Equation~\ref{eq:SinglePL}).

There does not appear to be a strong correlation between the empirical light curve fit parameters and their peak luminosities or rise times, at least not in this sample.
There is a tendency for larger values of $a_1$ to correspond to brighter peak luminosities in the \cite{2016Morozova} light curves, but this trend is very weak in the observed sample.
We see some of the expected correlation between the parameter $a_2$ and the light curve rise time in Figure~\ref{fig:PLParameters}, but the quantitative agreement is poor compared to the uncertainties, confirming the need for a more complex model if the fits are to be used to estimate the rise time.

\section{Comparison to Theoretical Models}
\label{sec:modeling}

Next we compare our \textit{TESS} observations to several theoretical models.
First, following \cite{2016Garnavich}, we use the semi-analytic treatments of \cite{2010Nakar} and \cite{2011Rabinak}.
Next, we examine the more detailed numerical simulations of \cite{2016Morozova}.
Finally, we investigate using the simulated light curves to calibrate the semi-analytic treatments.

\begin{figure}
\centering
\includegraphics[width=\columnwidth]{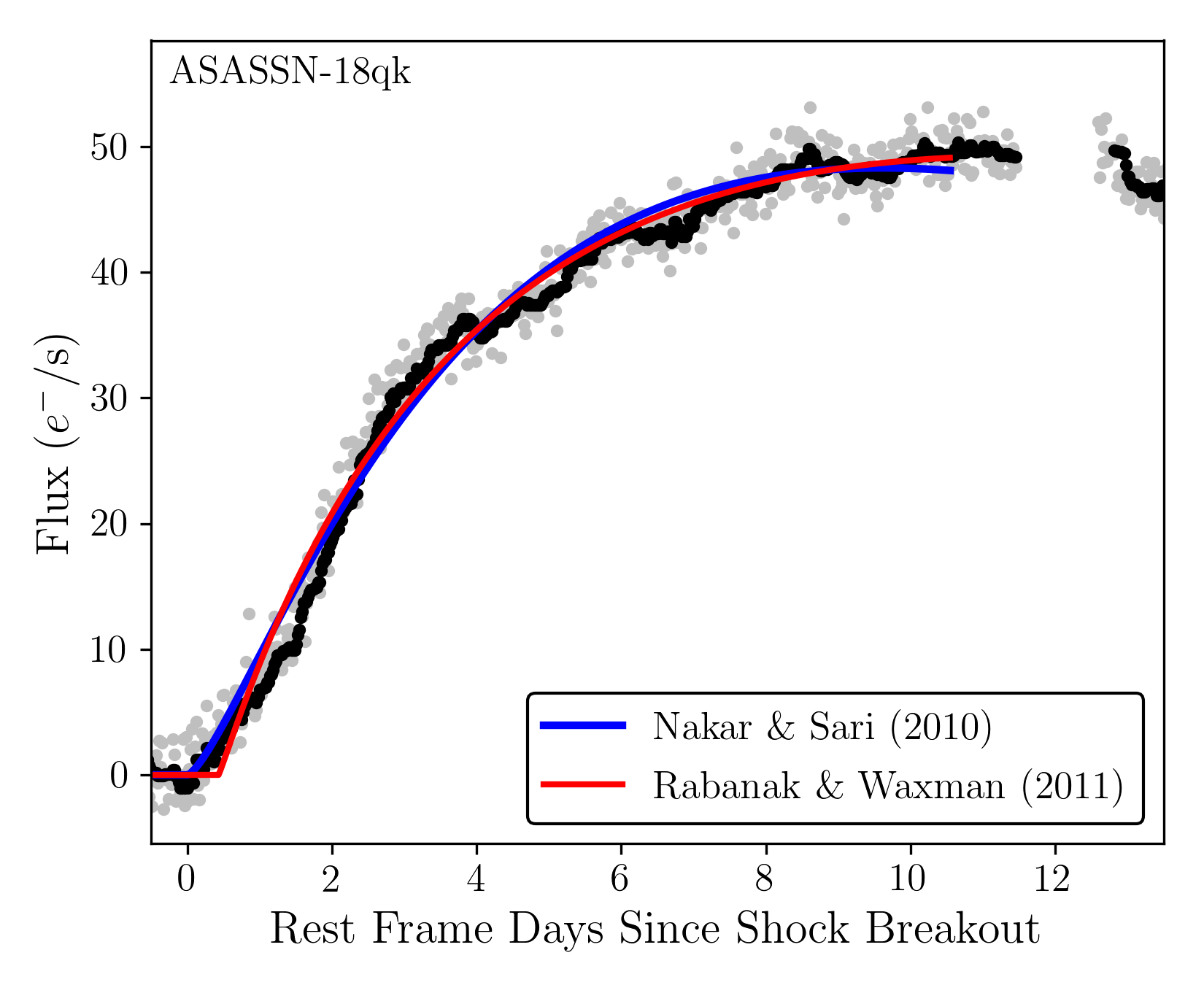}
\includegraphics[width=\columnwidth]{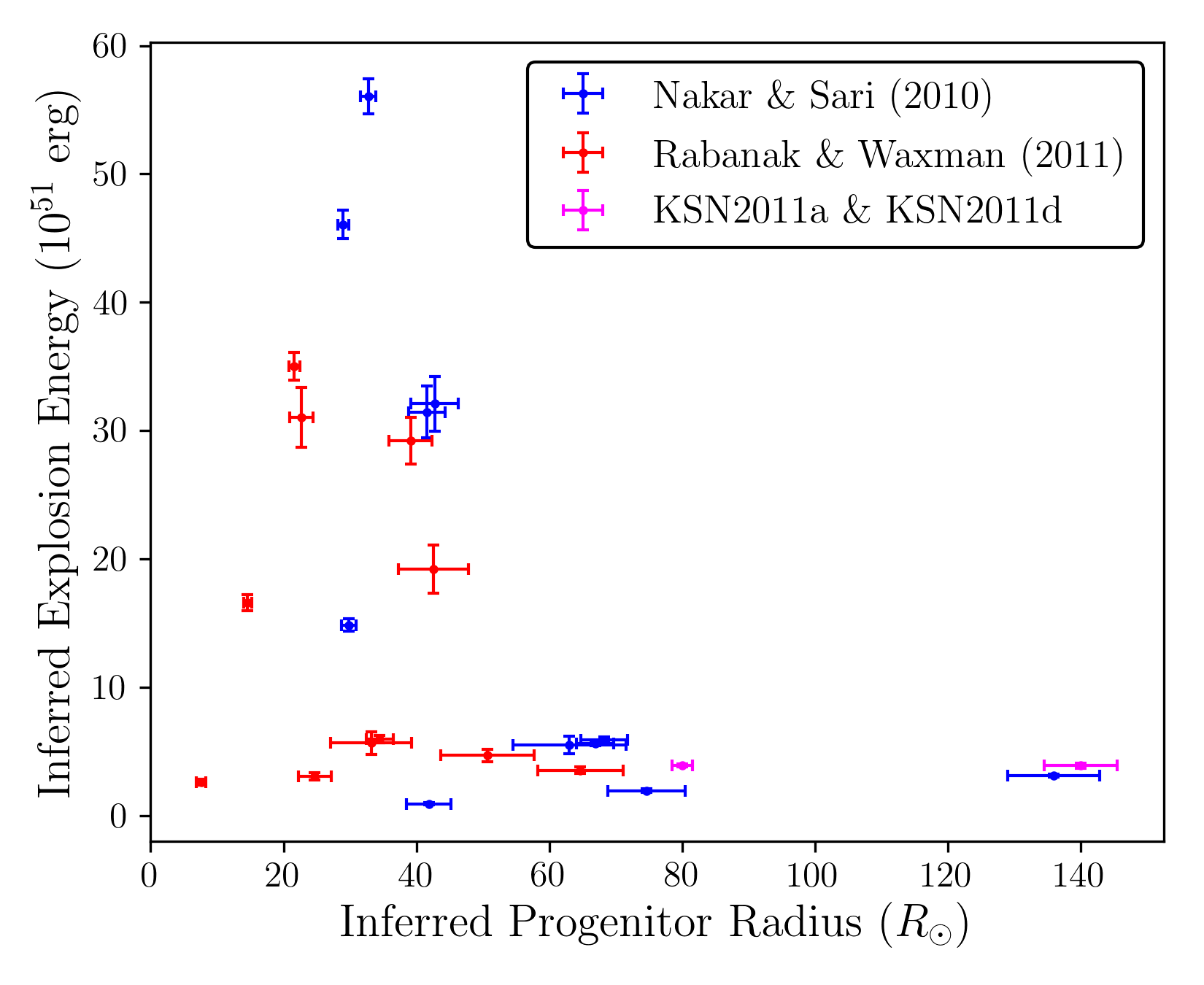}
\caption{Fits to \textit{TESS} data using the semi-analytic models of \protect\cite{2010Nakar} and \protect\cite{2011Rabinak}.
The top panel shows the best-fit models for ASASSN-18qk, a representative example of the fits obtained for the SNe~II sample.
After assuming a 12.5~$M_{\odot}$ progenitor, the best-fit \protect\cite{2010Nakar} model implies a progenitor radius of 29~$R_{\odot}$ and an explosion energy of $46\times10^{51}$~erg, while the best-fit \protect\cite{2011Rabinak} model implies a progenitor radius of 22~$R_{\odot}$ and an explosion energy of $35\times10^{51}$~erg.
Such implausible combinations of small radii and large explosion energies are a general feature of using these semi-analytic models to fit \textit{TESS} observations.
This can be seen in the lower panel, which shows the best-fit parameters for the 11 well-observed SNe~II in our sample. Updated fit parameters for KSN2011a and KSN2011d (shown in magenta) are also included in the lower panel.
}
\label{fig:RabinakNakarFits}
\end{figure}

\subsection{Semi-Analytic Models}
\label{subsec:analytic}

We first analyze our \textit{TESS} data using the same semi-analytic models that \cite{2016Garnavich} used when studying KSN2011a and KSN2011d.
The \cite{2011Rabinak} and \cite{2010Nakar} models describe core-collapse explosions as a time-dependent blackbody parameterized by the explosion energy, the density structure and opacity of the ejecta, and the mass and radius of the stellar progenitor.
If one makes some reasonable assumptions about the ejecta and assumes a fixed progenitor mass, the models depend primarily on the explosion energy and progenitor radius.
The model radius largely determines the light curve rise time, and the explosion energy largely determines the peak luminosity.
In principle, these models should be a good tool for estimating these generally inaccessible parameters of the explosion when coupled with high cadence \textit{TESS} or \textit{Kepler} light curves.

For our analysis we use semi-analytic models that assume a power-law density structure with an index of $n = 3/2$, appropriate for the efficiently convective envelopes of red supergiant stars (RSGs).
For the \cite{2010Nakar} model these are simply their Equations 29 and 31.
Some care must be taken when using the temperature description from \cite{2011Rabinak}, however, as the unmodified $T_{ph}$ from Equation 13 does not describe the observed blackbody.
This is instead described by the color temperature, $T_{col}$, and
for the relevant timescales, $T_{col}\approx1.23 T_{ph}$ (see \citeauthor{2011Rabinak}'s Figure~1).
Thus, for the \cite{2011Rabinak} model we use their Equations 13 and 14 and include the 1.23 scaling factor to convert $T_{ph}$ to $T_{col}$.
We then assume a fully ionized hydrogen envelope (with opacity $\kappa = 0.34 \textrm{cm}^2 \textrm{g}^{-1}$) and set the normalization of the ejecta density to $f_p = 0.1$, a value consistent with RSG progenitors \citep{2004Calzavara}.
We also assume a progenitor mass of 12.5 $M_\odot$, although we note that there is little difference if we instead use 15 $M_\odot$ like \cite{2016Garnavich}.

The results are shown in Figure~\ref{fig:RabinakNakarFits}.
While these models fit the \textit{TESS} data well, we find that they require implausibly large explosion energies and small progenitor radii.
For example, ASASSN-18qk is best fit with an explosion energy of $35\times10^{51}$ erg and a progenitor radius of 22~$R_{\odot}$ in the \cite{2011Rabinak} treatment and with an explosion energy of $46\times10^{51}$ erg and a progenitor radius of 29~$R_{\odot}$ in the \cite{2010Nakar} treatment.
These fits to ASASSN-18qk are shown in the top panel of Figure~\ref{fig:RabinakNakarFits}, and the best-fit values for the SNe~II sample are shown in the lower panel.
The explosion energies estimated from these fits are more than an order of magnitude larger than the typical $\sim10^{51}$~erg values expected for core-collapse events, and the radii are small compared to the observed $500-1000~R_\odot$ radii of RSGs (see, e.g., \citealt{2005Levesque}).

\cite{2017Sapir} provide a likely explanation for these results.
A fundamental assumption of these semi-analytic models is that the ejecta opacity remains constant over time, meaning that $T$ will depend primarily on $R$.
This assumption is reasonable for hydrogen dominated ejecta as long as $T \gtrsim 8100~\textrm{K}=T_R$.
At lower temperatures, recombination becomes important and modifies the opacity, complicating the relationship between $T$ and $R$ (See Figure~1 of \citealt{2017Sapir}).
At longer wavelengths ($\lambda_{lim} > hc/4k_BT_{R}\approx4400$~\AA{}) like those of the \textit{TESS} filter, the light curve peak occurs after recombination becomes significant and the assumption of constant opacity no longer holds.
Additionally, the longer rise times associated with redder filters exacerbate potential issues with shell curvature (Eq. 17 in \citealt{2011Rabinak}).
In practice this means that these models will become increasingly inaccurate for filters at wavelengths longer than that of the $U$-band (See Section 6.3 of \citealt{2017Sapir}).

While \textit{Kepler} observations use a shorter wavelength filter than \textit{TESS}, they are not particularly well-suited to this treatment either, as the \textit{Kepler} bandpass covers 4200--9000~\AA, peaking at 5750~\AA{} \citep{KeplerInstrumentHandbook}.
In their \cite{2011Rabinak} models of the \textit{Kepler} supernovae KSN2011a and KSN2011d, \cite{2016Garnavich} note that the radii of 280 $R_{\odot}$ and 490 $R_{\odot}$ they obtain are also small when compared to those observed for RSGs.
Their analysis differs from ours somewhat in that they assume $f_p=1.0$ and use $T_{ph}$ for their modeling.
Note that there is a typo in their text stating that the models were computed using a density parameter of $f_p=0.1$.
If we make the same assumptions as \cite{2016Garnavich}, we replicate their results.
When we re-fit the KSN2011a and KSN2011d data instead using $T_{col}$ and assuming $f_p=0.1$, we find that the explosion energy for both events increases from $2.0$ to $3.9\times10^{51}$ erg, and the progenitor radii are reduced by more than 70\%, to 80 $R_{\odot}$ and 140 $R_{\odot}$, respectively.
\cite{2016Morozova} also noted that fitting the semi-analytic \cite{2010Nakar} model to their synthetic light curves led to overestimates of the explosion energy and underestimates of the progenitor radius.
We will explore the question of whether these semi-analytic treatments can be empirically calibrated using the more detailed numerical models in Section~\ref{subsec:calibration}

\subsection{Numerical Simulations}
\label{subsec:numerical}

\begin{figure*}
    \centering
    \includegraphics[width=\textwidth]{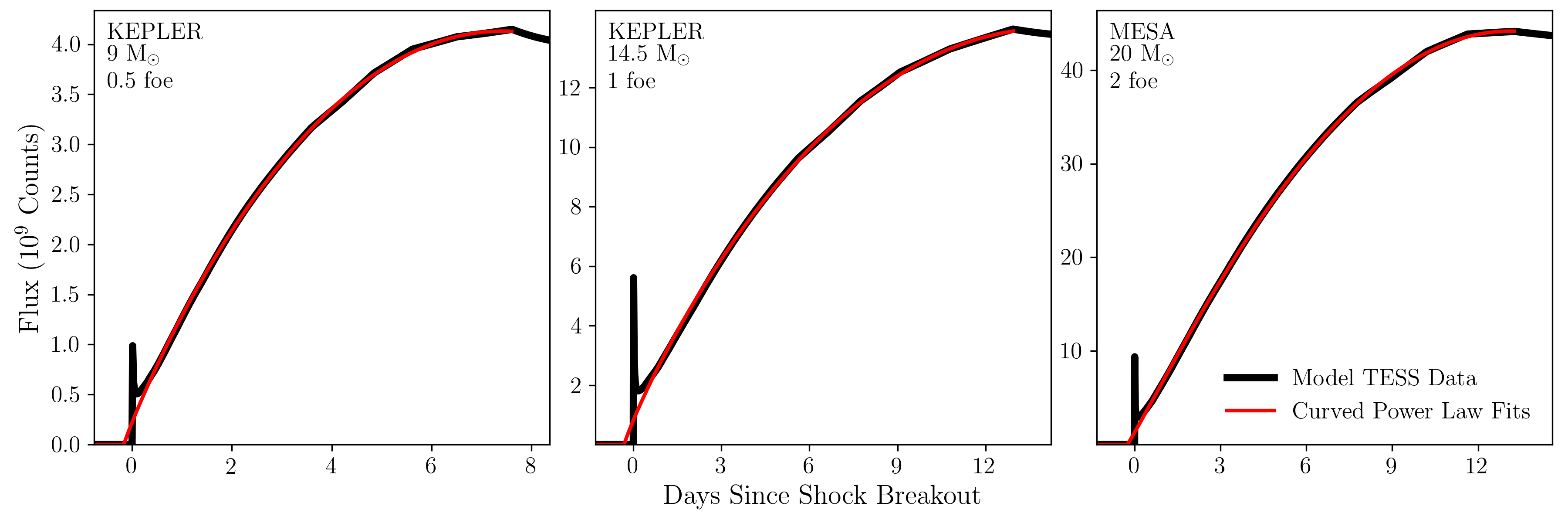}
    \caption{Examples of curved power-law fits to synthetic light curves calculated from the simulations of \protect\cite{2016Morozova}. The explosion energy and progenitor mass of the models increases from left to right. The left and middle panels show \textsc{KEPLER} progenitors, while the right panel shows a \textsc{MESA} progenitor. For each light curve the synthetic \textit{TESS} data is shown in black, and the best-fit curve is shown in red. The brief shock breakout spike at the start of the light curves is excluded from the fits.}
    \label{fig:ExampleModels}
\end{figure*}

\begin{figure*}
    \centering
    \includegraphics[width=\textwidth]{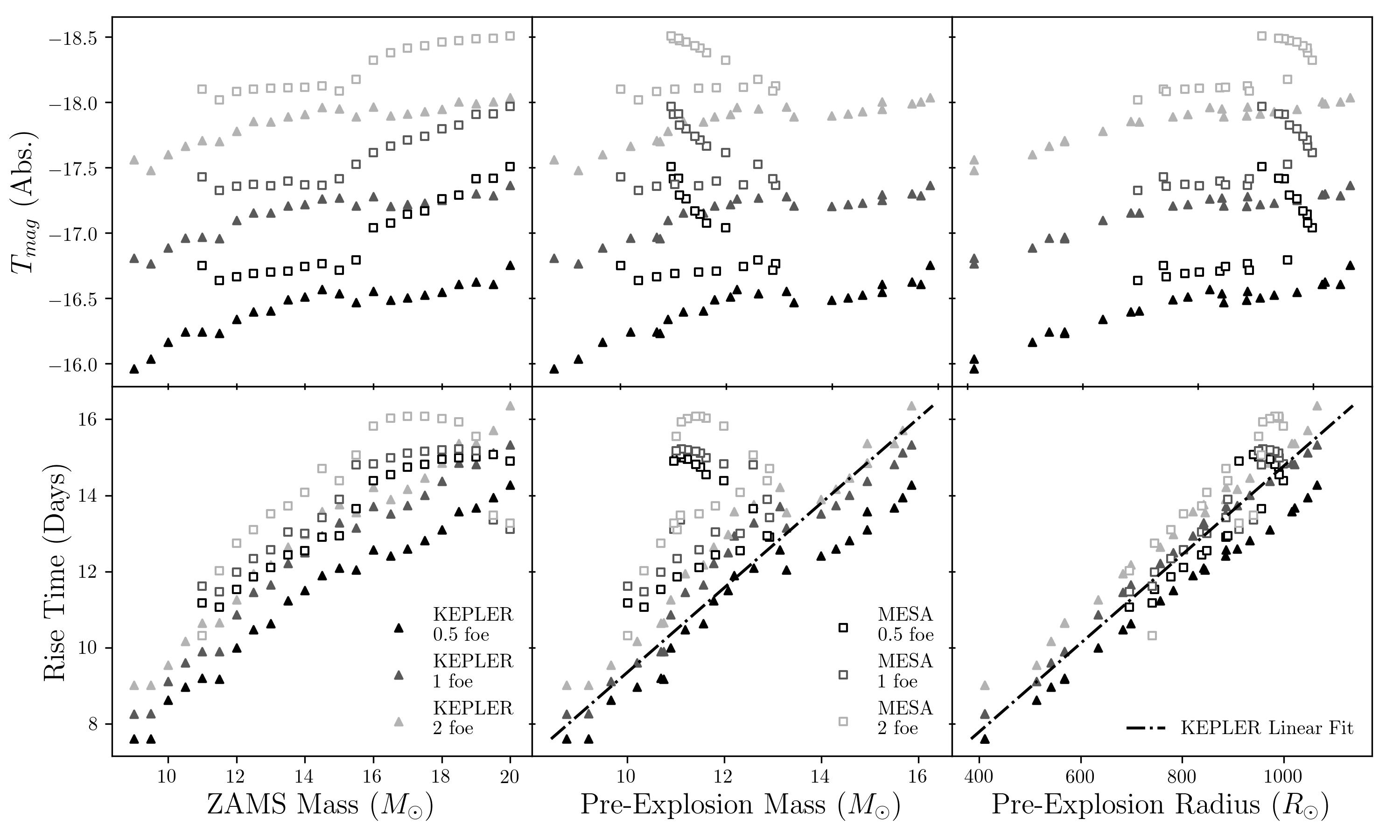}
    \caption{Peak absolute \textit{TESS} magnitude and rise time for the synthetic light curves computed by \protect\cite{2016Morozova} compared to the ZAMS mass, pre-explosion mass, and pre-explosion radius of the simulated progenitors. Progenitors simulated using \textsc{MESA} are shown as open squares, and those simulated using \textsc{KEPLER} are shown as solid triangles. Explosion energy is indicated by shading, with lighter shades corresponding to higher explosion energy. Note that a ``foe'' is equal to $10^{51}$ erg.}
    \label{fig:ModelParameters}
\end{figure*}

\begin{figure}
    \centering
    \includegraphics[width=\columnwidth]{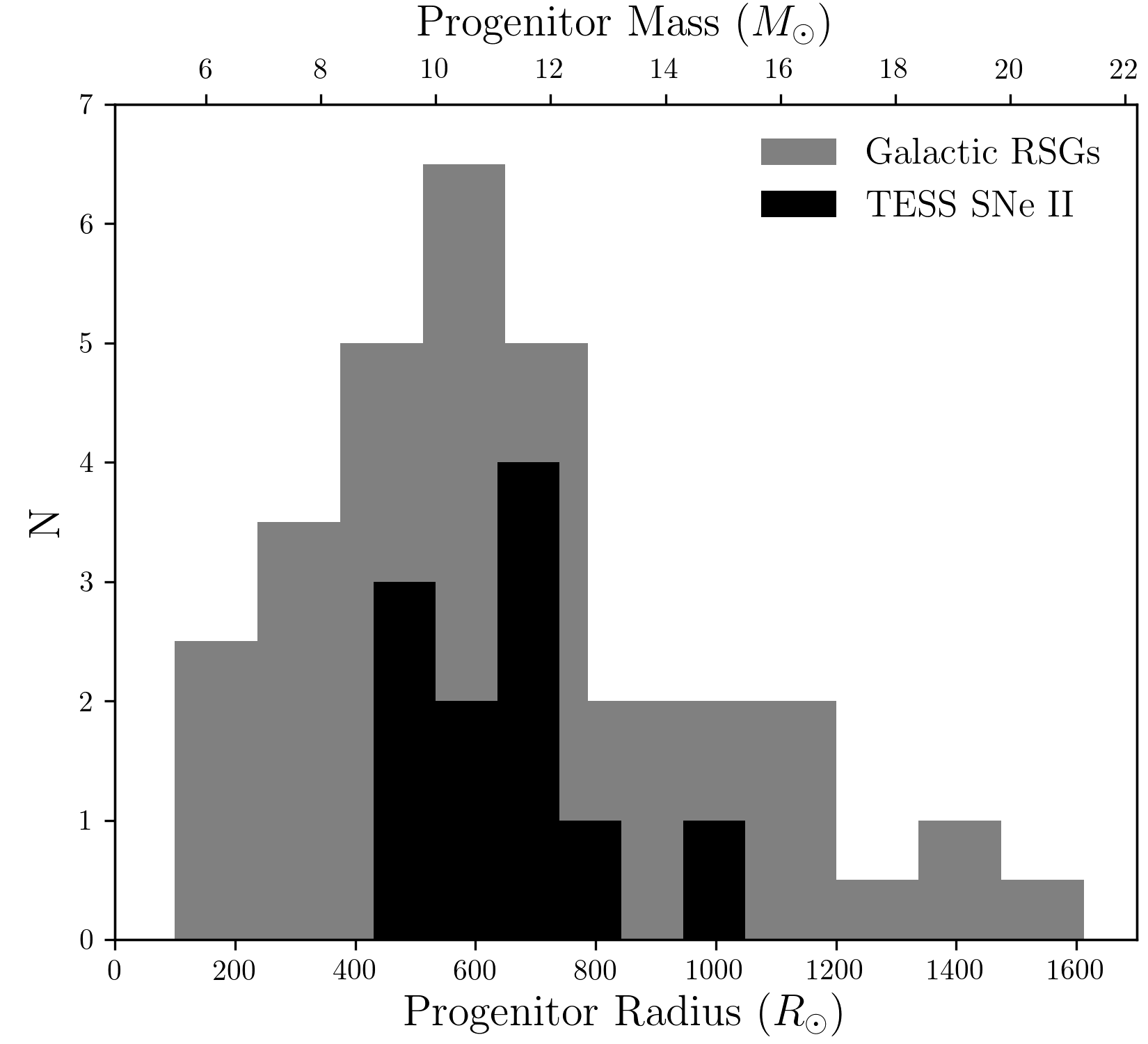}
    \caption{Comparison between the inferred radii of the SNe~II progenitors in our sample and the stellar radius estimates of the Galactic RSGs studied by \protect\cite{2005Levesque}. To ease comparison, the Galactic RSG bins have been scaled down by a factor of two. The two distributions are broadly similar, and both have median values around $R\sim640~R_\odot$, corresponding to $M\sim11~M_\odot$.}
    \label{fig:RadiusHistogram}
\end{figure}

Next, we compare our observations to the suite of 126 numerical SNe~IIP simulations presented by \cite{2016Morozova}.
\cite{2016Morozova} used the stellar evolution codes \textsc{MESA} \citep{Paxton2011,2013Paxton,2015Paxton} and \textsc{KEPLER} \citep{1978Weaver,2007Woosley,2014Sukhbold,2015Woosley,2016Sukhbold} to produce two sets of nonrotating, solar-metallicity red supergiant (RSG) stars to serve as progenitors.
To minimize confusion, we will refer to the stellar evolution code as \textsc{KEPLER} and the spacecraft as \textit{Kepler}.

There are 23 \textsc{KEPLER} models in total, ranging in zero-age main sequence (ZAMS) mass from 9 to 20 $M_\odot$, and 19 \textsc{MESA} models ranging in ZAMS mass from 11 to 20 $M_\odot$, both in increments of 0.5 $M_\odot$.
\cite{2016Morozova} explode these progenitor models using the SuperNova Explosion Code \citep[\textsc{SNEC};][]{2015Morozova}, which also generates both bolometric and filter-specific light curves.
Using what is often referred to as the ``thermal bomb'' mechanism, \textsc{SNEC} initiates the explosions by injecting energy into the inner regions of the progenitor model after excising the innermost 1.4 $M_\odot$ to simulate the newly formed compact object.
The amount of energy injected is chosen to yield the desired explosion energy.
For each of the 42 progenitor models this process is repeated three times to produce final (asymptotic) explosion energies of $E_{fin}=0.5$, $1.0$, and $2.0$ foe, where one foe is $10^{51}$ erg.

For the \textsc{MESA} models with ZAMS masses $\gtrsim15~M_\odot$, mass-loss is so significant that a larger ZAMS mass does not lead to larger pre-explosion mass.
For example, both the 20~$M_\odot$ and 12.5~$M_\odot$ ZAMS mass \textsc{MESA} progenitors have pre-explosion masses of $\sim11~M_\odot$.
This leads to the features seen in several of the Figure~\ref{fig:ModelParameters} panels, and is ultimately a product of the strong \cite{1988deJager} wind mass-loss prescription used for the \textsc{MESA} models.
This prescription is argued to over-estimate the mass-loss rate (e.g., \citealt{2014Smith}).
The \textsc{KEPLER} models use the \cite{1990Nieuwenhuijzen} mass-loss prescription, where this effect is only seen for ZAMS masses above $23~M_\odot$, beyond the range of masses considered by \cite{2016Morozova}.
We will focus primarily on comparisons to the \textsc{KEPLER} models in our discussion.

The \cite{2016Morozova} analysis focuses on synthetic $g$-band observations, but since \textsc{SNEC} models supernova emission as a blackbody, we can convert the bolometric luminosity and absolute $g$-band magnitudes from their work (all of which are available online at https://stellarcollapse.org/Morozova2016) into synthetic \textit{TESS} observations.
We use the \textsc{pysnphot} package \citep{2013pysynphot} to compute absolute $g$-band magnitudes and bolometric luminosities for a large sequence of $R_\odot$ blackbodies with 1 K spacing in $T_{eff}$ from 3,000 K to 303,000 K.
To match the \cite{2016Morozova} models we first use
\begin{equation}
    \log[L] = \log[L(T_{eff},R_\odot)]+2\log [R/R_\odot]
\end{equation}
to determine the blackbody radius $R$ necessary to match the bolometric luminosity for each $R_\odot$ blackbody in our grid.
We then use
\begin{equation}
    M_g = M_g(T_{eff},R_\odot) - 5 \log [R/R_\odot]
\end{equation}
to determine which combination of $T_{eff}$ and $R$ best matches the \textsc{SNEC} data, and this process is repeated for each epoch in the synthetic light curves.
Three representative examples of these synthetic $TESS$ observations are shown in Figure~\ref{fig:ExampleModels}, and $TESS$-band versions of all 126 model light curves from \cite{2016Morozova} are included in the online supplementary material.

We fit the curved power-laws to the synthetic data after excluding the shock breakout feature, as \textsc{SNEC} light curves are unreliable for the first $\sim1/4$ days before the photosphere recedes and becomes better resolved \citep{2015Morozova}.
These fits are shown by the red curves in Figure~\ref{fig:ExampleModels}, and as was found for the observed light curves, Eqn.~\ref{eq:CurvedPL} fits the rising phases of the model light curves very well.
The \textit{TESS}-band peak absolute magnitude and rise time for all of the \cite{2016Morozova} models are shown in Figure~\ref{fig:ModelParameters} as a function of their ZAMS mass, pre-explosion mass, and pre-explosion radius.

To first order, these synthetic \textit{TESS} light curves are comparable to the observed sample.
Like the observed sample, the model light curves have rise times of order two weeks and peak absolute \textit{TESS}-band magnitudes of approximately $-18$ mag.
In detail, however, the two samples differ somewhat.
Notably, the \cite{2016Morozova} light curves cluster in relatively confined portions of parameter space compared to the significant diversity exhibited by the observed SNe.
This is particularly evident in the $\alpha$ and $a_2$ distributions shown in Figure~\ref{fig:SingleCompParameters} and Figure~\ref{fig:PLParameters}, respectively.
There are also differences in the peak brightness, as even the brightest model light curves only attain absolute \textit{TESS}-magnitudes of about $-18.5$ mag, a mark exceeded by several SNe~II in our sample.

Examining the relationships shown in Figure~\ref{fig:ModelParameters}, a number of trends are readily apparent.
First, larger explosion energies and larger progenitors tend to produce slower rising, more luminous light curves, with the peak luminosity being driven primarily by the explosion energy and rise time being driven primarily by the progenitor size, just as in the semi-analytic models.
Among the \textsc{MESA} models there is a moderately strong correlation between ZAMS mass and rise time, but the relationship between pre-explosion mass and rise time is complicated by the effects of mass loss.
For the \textsc{KEPLER} models there is a tight linear correlation between rise time and both ZAMS mass and pre-explosion mass.
\cite{2016Morozova} found that there is a strong correlation between $g$-band rise time and progenitor radius.
The rise times are naturally longer for the redder $TESS$-band, but the correlations between the rise time and the mass or radius are retained.
While we focus our subsequent fits on the \textsc{KEPLER} models, the \textsc{MESA} models follow essentially the same radius-rise time relation.

The tight relation between progenitor radius and rise time for the \textsc{KEPLER} models shown in Figure~\ref{fig:ModelParameters} is well-fit by
\begin{equation}
    R/R_\odot = 85.8 \cdot \big( t_{rise}/\textrm{days} \big) - 267.6,
\end{equation}
with a scatter of $\sigma_R=58.9~R_\odot$ in radius.
We obtain a similar fit for the \textsc{KEPLER} models between progenitor mass and rise time of
\begin{equation}
    M/M_\odot = 0.897 \cdot \big( t_{rise}/\textrm{days} \big) + 1.61,
\end{equation}
with a scatter of $\sigma_M=0.77~M_\odot$ in mass.

Figure~\ref{fig:RadiusHistogram} shows a histogram of the  radius estimates for the twelve SNe~II in our sample, using the above relations to convert their measured rise times to estimates of their progenitor radii and masses.
For comparison, we also include a histogram of the 62 Galactic RSGs for which \cite{2005Levesque} were able to obtain radius estimates from MARCS stellar atmosphere models \citep{2003Plez,2003Gustafsson}.
The two distributions are broadly similar, each with median values near $R\sim640~R_\odot$ (corresponding to $M\sim11~M_\odot$).
It is worth noting that this result stands in some contrast to the Sloan Digital Sky Survey and Supernova Legacy Survey SNe~II sample of \cite{2015GonzalezGaitan}, which favors progenitors smaller than $500~R_\odot$.

\subsection{Semi-Analytic Model Calibration}
\label{subsec:calibration}

\begin{figure*}
    \centering
    \includegraphics[width=\textwidth]{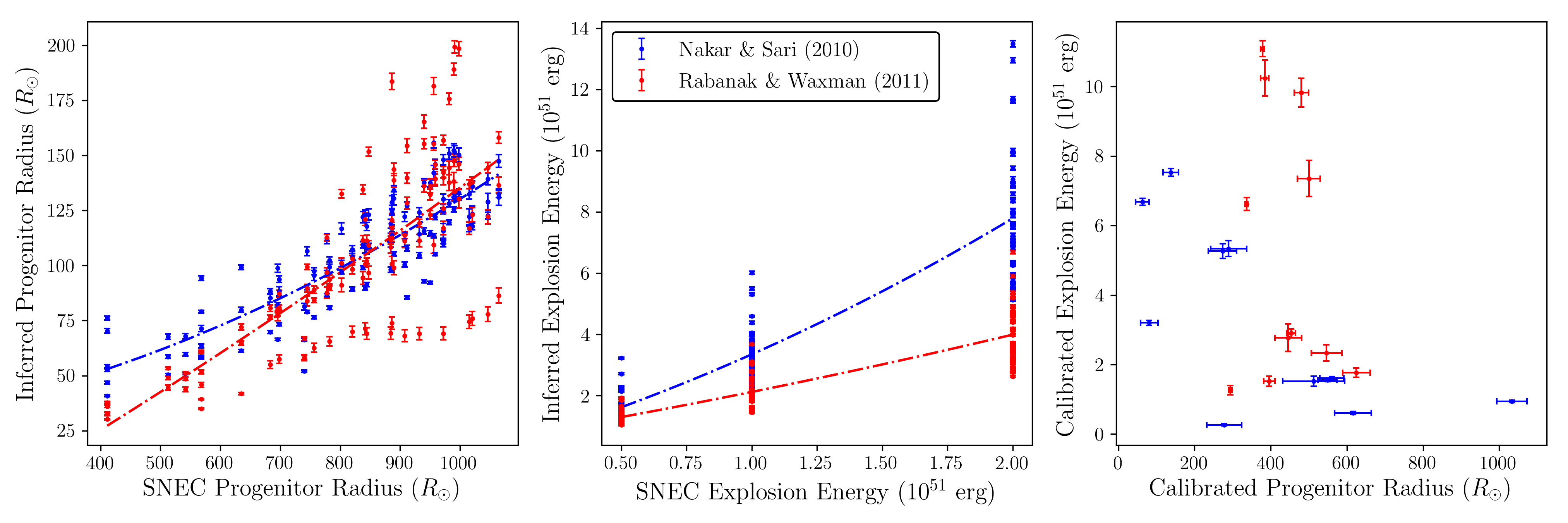}
    \caption{The semi-analytic models of \protect\cite{2010Nakar} and \protect\cite{2011Rabinak} calibrated using the known progenitor properties of the \protect\cite{2016Morozova} \textsc{SNEC} models. The left and center panels show the model parameters obtained when fitting the rising light curves of the \protect\cite{2016Morozova} light curves using the \protect\cite{2010Nakar} and \protect\cite{2011Rabinak} treatments. The semi-analytic fits shown here assume a progenitor mass of 12.5~$M_{\odot}$. The left panel shows the progenitor radii inferred by the semi-analytic models as a function of the actual \textsc{SNEC} progenitor radius, and the center panel shows the inferred explosion energy as a function of its true value. Blue markers indicate the \protect\cite{2010Nakar} values, and red markers indicate the \protect\cite{2011Rabinak} values. Best-fit quadratics to each set of models are shown by the dot-dashed lines. The right panel shows the best-fit parameters for the 11 well-observed SNe~II in our sample after calibrating their semi-analytic fit parameters using the best-fit relations for progenitor radius and explosion energy. This procedure leads to more physically reasonable values, although the inferred explosion energies are still quite high.}
    \label{fig:CalibratedParams}
\end{figure*}

While the physical values inferred from the \cite{2010Nakar} and \cite{2011Rabinak} models are implausible, the fits themselves are quite good (e.g., the top panel of Figure~\ref{fig:RabinakNakarFits}).
Here we investigate the possibility of calibrating the semi-analytic treatments using the models from \cite{2016Morozova}.
To do this we fit the rising light curves of all 126 \cite{2016Morozova} models using the \cite{2010Nakar} and \cite{2011Rabinak} treatments, adopting the same fixed 12.5$M_\odot$ progenitor mass as in Section~\ref{subsec:analytic} to facilitate consistent comparisons.
Repeating this procedure using the true \textsc{SNEC} progenitor mass instead of a fixed value has no significant effect on the results because the semi-analytic models are only weakly dependent on progenitor mass.

The left and center panels of Figure~\ref{fig:CalibratedParams} show the progenitor radii and explosion energies inferred by the semi-analytic treatments as a function of the true values in the \cite{2016Morozova} \textsc{SNEC} simulations.
Consistent with our earlier results and the discussion in \cite{2016Morozova}, the \cite{2010Nakar} and \cite{2011Rabinak} fits underestimate the true radii and overestimate the true explosion energies by considerable margins.
They are, however, reasonably well-correlated.

This allows us to obtain best-fit quadratic curves as a means of calibrating the semi-analytic results.
These best-fit curves are shown as the dot-dashed lines in Figure~\ref{fig:CalibratedParams}.
For the \cite{2010Nakar} treatment, the best-fit calibration curves are
\begin{equation}
    R_{NS10} = 26.3 + 0.0375\cdot R_{SNEC} + 6.64\times10^{-5}\cdot R_{SNEC}^2,
\end{equation}
and
\begin{equation}
    E_{NS10} = 0.224 + 2.47\cdot E_{SNEC} + 0.655\cdot E_{SNEC}^2.
\end{equation}
In these expressions, radii are in units of $R_\odot$ and explosion energies are given in units of $10^{51}$ erg.
The residual scatters about the fits are $\sigma_R=13.4~R_{\odot}$ and $\sigma_E=1.3\times10^{51}$~erg,
respectively.  
For the \cite{2011Rabinak} treatment, the best-fit calibration curves are
\begin{equation}
    R_{RW11} = -39.1 + 0.152\cdot R_{SNEC} + 2.23\times10^{-5}\cdot R_{SNEC}^2,
\end{equation}
and
\begin{equation}
    E_{RW11} =  0.559 + 1.41\cdot E_{SNEC} + 0.153\cdot E_{SNEC}^2.
\end{equation}
The scatters about these fits are $\sigma_R=24.8~R_{\odot}$ and $\sigma_E=0.6\times10^{51}$~erg.

We then use these fits curves to calibrate the inferred progenitor radii and explosion energy estimates obtained from the \cite{2010Nakar} and \cite{2011Rabinak} models.
Doing so we obtain the right panel of Figure~\ref{fig:CalibratedParams}.
Compared to the non-calibrated model parameters shown in the lower panel of Figure~\ref{fig:RabinakNakarFits}, the calibrated parameters are considerably more reasonable, 
particularly the radius estimates.
While improved from the non-calibrated explosion energy estimates (which are in many cases larger than $30$ or even $50\times10^{51}$~erg), the explosion energy estimates remain about an order of magnitude larger than expected values, even after calibration.
The discrepancy in explosion energy may be due to the limited range of explosion energies used by the \cite{2016Morozova} models.
This forces significant extrapolation of the calibration curve in order to match the high luminosity observed events.

\section{Shock Breakout Signatures}
\label{sec:shockbreakout}

\begin{figure}
    \centering
    \includegraphics[width=\columnwidth]{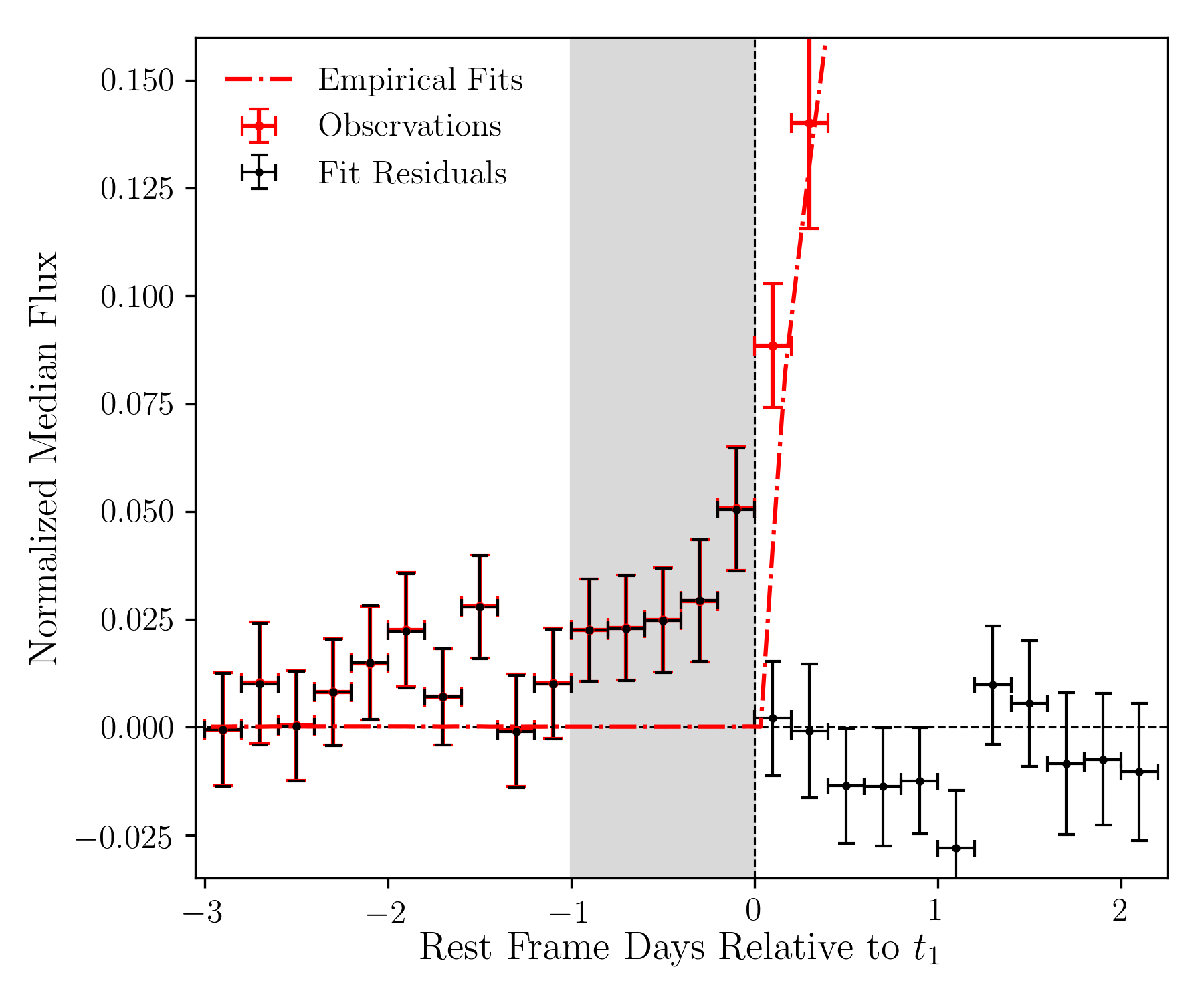}
    \caption{Stacked light curves of the 11 non-CSM interacting SNe~II in our sample near the time of first light. Residuals, shown in black, are calculated by subtracting the best-fit curved power-laws from the observed light curves, shown in red. Horizontal error bars represent the time interval for each bin, and vertical error bars show the 1$\sigma$ uncertainty in the median for the stacked fluxes and residuals from each bin. There is a clear, statistically significant excess visible in the stacked observations clustered just prior to $t=t_1$, the same time where we would expect to find signatures of shock breakout. We find a $3\sigma$ excess at $t-t_1=-0.1$ days, and this grows to $>5\sigma$ significance when we include all of the bins from $t_1-1.0$ to $t_1$ days (the shaded region shown above).}
    \label{fig:StackLC}
\end{figure}

\begin{figure}
    \centering
    \includegraphics[width=\columnwidth]{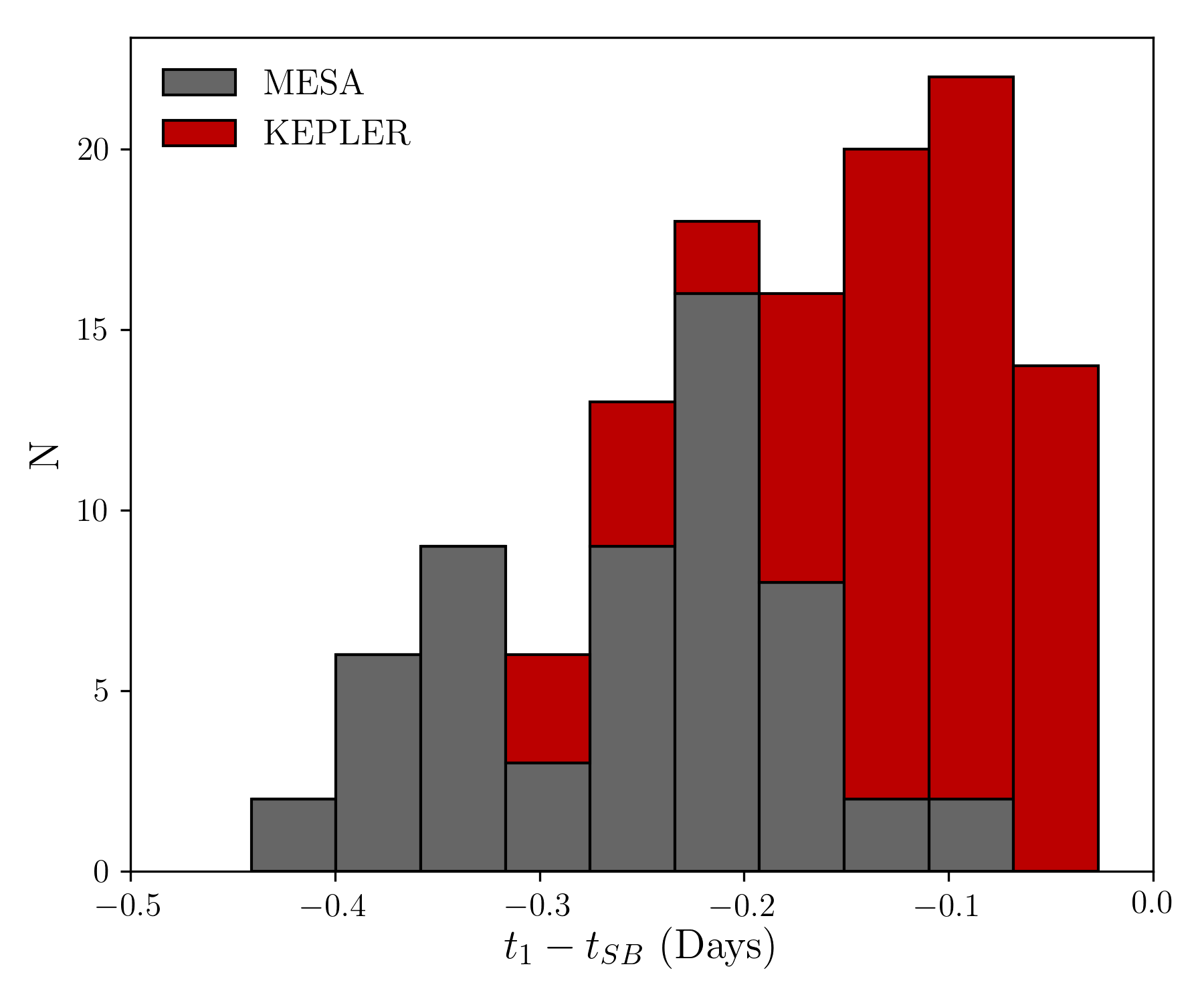}
    \caption{Histograms of the offsets between the inferred time of first light ($t_1$) and the time of shock breakout ($t_{SB}$) for the fits to the 126 synthetic light $TESS$ light curves calculated from the \protect\cite{2016Morozova} models. The \textsc{KEPLER} models are shown in scarlet, and \textsc{MESA} models are shown in gray.  Such offsets will contribute to temporally smearing the shock break out emission in Figure~\ref{fig:StackLC}.}
    \label{fig:Model_t1s}
\end{figure}

One notable feature of the \cite{2016Morozova} models discussed in Section~\ref{subsec:numerical} that we do not find in the observed sample is the unmistakable shock breakout spikes at the start of the supernova light curves.
The strength of this signal varies among the synthetic light curves, but it typically reaches $\gtrsim30\%$ of the peak brightness (see Figure~\ref{fig:ExampleModels}).
The time-steps used in the publicly available light curve files from \cite{2016Morozova} are just under 30 minutes, almost identical to the exposure length for a \textit{TESS} FFI, so we would expect to see a signal of similar strength in the observations.
However, visual inspection of the supernovae light curves in Figure~\ref{fig:AllSNe} shows that this is not the case.

The existence of shock-breakout emission is a robust prediction of core-collapse theory that has been detected previously in observations at shorter wavelengths, so it is very likely present in the \textit{TESS} data and is merely weaker than the \textsc{SNEC} models predict.
That \textsc{SNEC} would simulate shock breakout imperfectly is not surprising.
\cite{2015Morozova} note that during shock breakout the photosphere is in the outermost grid cell of the simulation and is spatially poorly resolved, rendering light curves in this early phase unreliable.
They also note that the code assumes local thermodynamical equilibrium (LTE), imposing the same temperature for radiation and matter, an invalid approximation during shock breakout.

Shock breakout occurs on a timescale comparable to the progenitor's light crossing time, $t_{lc}=R_*/c$.
For the $\sim640~R_\odot$ progenitors in our sample, this corresponds to a duration of about 30 minutes.
Because of this short timescale, averaging over the FFIs may not aid in the detection of the shock breakout peak for an individual SN because it smears the peak out.
We can, however, look for excess emission near the inferred time of first light, $t_1$ averaged over multiple SNe.

To do this, we stack the fit residuals for the 11 non-CSM interacting SNe~II observed by \textit{TESS} prior to explosion, subtract the best-fit curved power-laws from their observed light curves and combine all of the residuals.
The CSM interacting SNe~IIn are excluded from this analysis because the shock breakout signature is expected to be reprocessed and diluted by the dense CSM surrounding these supernovae.
Figure~\ref{fig:StackLC} shows the resulting median residuals, focused on the timescales where we would expect to find shock-breakout emission and normalized such that the peak flux of the light curve is unity.
The horizontal error bars indicate the time interval for each bin, and the vertical error bars show the $1\sigma$ uncertainty in the median for the stacked residuals from each bin.
There is a readily apparent excess visible in the residuals, peaking at $3\sigma$ significance just prior to $t_1$, the time when the power law rise begins.
The cumulative significance of the excess feature reaches $5\sigma$ when we incorporate all of the residuals in the 1 day prior to $t_1$, the shaded region shown in Figure~\ref{fig:StackLC}.
It is possible that this excess is due to other factors, but it is plausibly associated with shock break out, so we will proceed under this assumption.

The emission clearly is not a sharp spike, and we would not expect to see one in these stacked residuals.
The first issue is that the signals from the individual SNe will be in misaligned in time by any errors in the estimate of the rise time $t_1$. 
The measurement errors in $t_1$ are typically $0.25$~days (Tab.~\ref{tab:params}), but there are also probably systematic errors.
To explore this question, we show the distribution of time differences between our fits to the \cite{2016Morozova} models for $t_1$ and the actual time of shock breakout in Fig.~\ref{fig:Model_t1s}.
This suggests that we might expect additional systematic shifts
of $0.1$-$0.4$~days.
So in the stacked residuals, the signal will likely be smeared over $\sim 0.5$~days just by the uncertainties in how to temporally align the events.

The second issue is that while the shock break out peak lasts only approximately the light crossing time $t_{lc}$, it is followed by a slower phase in which the ejecta expand and cool nearly adiabatically before the rise to peak begins.
The time scale for this phase of $0.5$ to $1.0$ days for RSGs is comparable to the transition time scale between the planar and spherical phases of the \cite{2010Nakar} model.
While dimmer than the initial spike, this tail is not negligible and will temporally broaden
the signal.
Combining these two effects, the structure of the residuals does not seem surprising.

We can use the amplitude of the excess to roughly estimate (or limit) the amplitude of shock break out signals in the TESS band.
The energy of the excess is $E_{ex} \simeq \epsilon F_{peak} \Delta t$, where $\epsilon \simeq 0.02$ is the amplitude of the excess relative to the peak flux $F_{peak}$ (unity in Fig.~\ref{fig:StackLC}) and $\Delta t \simeq 0.5$~days is the duration of the observed excess.
This must be equal to the energy in the break out pulses, $F_{SBO} \cdot t_{SBO}$, where $F_{SBO}$ is the mean flux over time $t_{SBO}$.
Combining these, we must have that $F_{SBO}/F_{peak} \simeq \epsilon \Delta t/t_{SBO}$. 
Clearly we cannot have that most of the shock break out energy is emitted in the light crossing time, since for $t_{SBO} \simeq t_{lc} \simeq 0.5$~hours, $F_{SBO}/F_{peak} \simeq 0.5$ would produce signals easily visible in Fig.~\ref{fig:AllSNe}.
However, if we spread the emission over the overall time scale of the initial decline, $t_{SBO} \simeq 8$~hours, then $F_{SBO}/F_{peak} \simeq 0.06$.
Such a smaller amplitude signal would be relatively easy to hide for the present sample.

\section{Discussion and Conclusions}
\label{sec:conclusions}

In this work we have presented the first \textit{TESS} observations of core-collapse supernovae.
Due to its large survey area and continuous monitoring, \textit{TESS} is particularly well-suited for obtaining high-cadence early-time observations of bright extragalactic transients such as these SNe.
However, aspects of the \textit{TESS} images like their large pixel size, the straps, and the many scattered light artifacts can make analyzing these observations difficult.
In Section~\ref{sec:LIGER} we have described an image subtraction pipeline that addresses the most common issues present in \textit{TESS} data.
We have optimized this pipeline for the study of extragalactic transients, but these techniques would likely be beneficial for other \textit{TESS} applications as well.

We do not identify any strong trends between the parameters of our empirical light curve fits and the peak luminosities of the SNe.
The semi-analytic models of \cite{2010Nakar} and \cite{2011Rabinak} fit the data well, but the resulting estimates of the explosion energies and progenitor radii are not physical, probably because the \textit{TESS} bandpass is too red.
This also appears to be true for \textit{Kepler} observations.
Numerical light curves computed using \textsc{SNEC} yield more plausible estimates, and may provide a means through which the semi-analytic models can be calibrated.
We briefly explored this possibility in Section~\ref{subsec:calibration}.
Here we simply fit polynomials to convert the semi-analytic energies and radii to better agree with the input models.
Doing so produced more physically reasonable radius estimates and improved explosion energies, although even after calibration the inferred explosion energies remained quite high.
A similar approach, which we did not explore here, would be to modify the dimensionless factors and perhaps the exponents of the semi-analytic models to achieve the same end.

Broadly speaking, the synthetic \textit{TESS} light curves produced from the \cite{2016Morozova} models are comparable to what we observe in our sample.
The empirical fit parameters for the two data sets are reasonably similar, the rise times are of order two weeks, and the peak absolute \textit{TESS}-band magnitudes are of order $-18$ mag.
In detail, however, the \cite{2016Morozova} models appear to differ somewhat from the observed light curves.
The \textsc{SNEC} models cluster in relatively confined portions of parameter space compared to the observed light curves, failing to reproduce the full diversity implied by the observations.
The observed power law indices found for the rise tend to be shallower but also more diverse.
Additionally, even the most energetic \textsc{SNEC} explosions of the most massive model progenitors only reach absolute \textit{TESS}-magnitudes of about $-18.5$ mag. 
Four of the twelve SNe~II in our sample have higher peak luminosities.

A detailed study of these discrepancies is beyond the scope of this work, but we can consider potential explanations.
When compared to other numerical models of core-collapse explosions, the \cite{2016Morozova} models are relatively simple.
They do not include contributions from radioactive $^{56}$Ni or interactions with CSM, for example.
These simplifications allow \cite{2016Morozova} to study a range of models broadly dispersed throughout the progenitor size and explosion energy parameter spaces.
In the absence of significant mixing, contributions from radioactive $^{56}$Ni are likely small during the early rise of core-collapse light curves, but subsequent studies by \cite{2017Morozova} and \cite{2018Morozova} have argued that emission enhancement due to CSM interaction is an extremely important aspect of early SNe~II light curves.
This is, however, inconsistent with pre-supernova observations of Type~II progenitors \cite{2018Johnson}.
A more likely explanation is that SN progenitors do not have the ``sharp'' edges of stellar evolution models, instead having extensions to their envelopes driven by pulsations, which do not in turn produce high mass density winds.
Additional problems may arise from simplifications in the \textsc{SNEC} treatment, like its assumption of LTE throughout the model.
\cite{2018Morozova} note that while \textsc{SNEC}'s bolometric light curves generally agree quite well with the multigroup radiation-hydrodynamic code \textsc{STELLA} \citep{1993Blinnikov,2011Blinnikov}, their synthetic filter light curves do exhibit minor discrepancies.  More self-consistent explosion models, like those of the PUSH framework \citep{2015Perego,2020Curtis}, might also modify the very early light curves.

We find a mean pre-rise flux excess for the Type~II SNe of $\sim 2\%$ of the peak flux in the $\sim 0.5$~days before the estimated start of the rising light curve which is plausibly due to the shock break out.
In this scenario, the excess cannot be dominated by energy from an initial peak lasting only the light crossing time ($\sim 0.5$~hours), as this would lead to visible peaks in the individual light curve.
The energy would have to be emitted over a longer time period.
We roughly estimate that the signal should be directly detectable for an SNe~II with a peak \textit{TESS} magnitude brighter than about 15~mag.
The switch from a 30-minute cadence to a 10-minute cadence in the extended \textit{TESS} mission will make searches for such emission from stripped SNe more feasible.
In addition to the bright SN~Ibn and SN~IIb presented here, \textit{TESS} has already observed multiple SNe~Ia \citep{2019Fausnaugh} and a tidal disruption event \citep{2019Holoien}  brighter than 15~mag, so  it is very likely that \textit{TESS} will observe a sufficiently bright SN~II to allow a direct detection.
Even without a direct detection, stacking analyses like that used here will steadily improve.

\textit{TESS} provides a valuable new means of studying core-collapse supernovae, one that will only become more significant as the sample of early-time \textit{TESS} observations grows. Better models
to interpret the early-time emission are clearly needed.  In particular, (semi-analytic) models of both the shock break out peak, its decay and the initial rise appropriate for these redder bands would be 
very useful, as would methods of using expensive numerical simulations to calibrate simpler models that can be easily fit to the data.
With larger numbers of SNe, we would also hope to see clear statistical patterns begin to appear in the distributions and correlations of the parameters describing the initial light curve rises. As a continuing mission, there remains the chance of a spectacularly bright SN that can be followed in detail at the full TESS FFI cadence.

\section*{Acknowledgements}

We greatly appreciate Viktoriya Morozova sharing model light curves with us, as well as Peter Garnavich's  willingness to discuss semi-analytic modeling approaches and share some of the fitting code used in \cite{2016Garnavich}.
We would also like to thank Tuguldur Sukhbold, Eli Waxman, and Sanjana Curtis for valuable discussions.
PJV is supported by the National Science Foundation Graduate Research Fellowship Program Under Grant No. DGE-1343012.
CSK, KZS, and BJS are supported by NSF grant AST-1907570.
CSK and KZS are also supported by NSF grant AST-181440.
BJS is also supported by NASA grant 80NSSC19K1717 and NSF grants AST-1920392 and AST-1911074.

This paper includes data collected by the \textit{TESS} mission, which are publicly available from the Mikulski Archive for Space Telescopes (MAST).
Funding for the \textit{TESS} mission is provided by NASA's Science Mission directorate.
We thank Ethan Kruse for continuing to upload animations of the TESS FFIs to YouTube, as these videos have been invaluable for investigating the systematics in our data.
In addition to the software cited in the main body of the paper we have also made use of \textsc{NumPy} \citep{2020NumPy}, \textsc{SciPy} \citep{2020SciPy}, \textsc{Astropy} \citep{2013Astropy}, \textsc{PyRAF} \citep{2012PyRAF}, \textsc{IPython} \citep{2007iPython}, \textsc{Matplotlib} \citep{2007Matplotlib}, \textsc{pandas} \citep{2010Pandas}, and \textsc{SAOImage DS9} \citep{2003ds9}

This work is heavily reliant on the ongoing All-Sky Automated Survey for Supernovae.
We thank the Las Cumbres Observatory and its staff for its continuing support of the ASAS-SN project.
ASAS-SN is supported by the Gordon and Betty Moore Foundation through grant GBMF5490 to the Ohio State University, and NSF grants AST-1515927 and AST-1908570.
Development of ASAS-SN has been supported by NSF grant AST-0908816, the Mt. Cuba Astronomical Foundation, the Center for Cosmology and AstroParticle Physics at the Ohio State University, the Chinese Academy of Sciences South America Center for Astronomy (CAS- SACA), and the Villum Foundation. 

In this work we have made extensive use the ZTF alert broker Lasair \citep{2019Lasair} and the ZTF data it provides access to.
Lasair is supported by the UKRI Science and Technology Facilities Council and is a collaboration between the University of Edinburgh (grant ST/N002512/1) and Queen's University Belfast (grant ST/N002520/1) within the LSST:UK Science Consortium.
ZTF is supported by National Science Foundation grant AST-1440341 and a collaboration including Caltech, IPAC, the Weizmann Institute for Science, the Oskar Klein Center at Stockholm University, the University of Maryland, the University of Washington, Deutsches Elektronen-Synchrotron and Humboldt University, Los Alamos National Laboratories, the TANGO Consortium of Taiwan, the University of Wisconsin at Milwaukee, and Lawrence Berkeley National Laboratories.
Operations are conducted by COO, IPAC, and UW.
This research has made use of ``Aladin sky atlas'' developed at CDS, Strasbourg Observatory, France 2000A\&AS..143...33B and 2014ASPC..485..277B.




\bibliographystyle{mnras}
\bibliography{TESS_CCSNe.bib} 

\bsp	
\label{lastpage}
\end{document}